\documentclass[trackchanges, twocolumn]{aastex701}
\usepackage{amsmath}
\usepackage{booktabs}
\usepackage{bm}

\providecommand{\apjs}{Astrophys. J. Suppl. Ser.}

\providecommand{\apjl}{Astrophys. J. Lett.}
\providecommand{\aap}{Astron. Astrophys.}

\begin{document}

\title{Beyond the Inner Slope $\gamma$: A Schwarzschild-based Probe of the Core-Cusp Problem in the Milky Way Dwarf Spheroidal Galaxies}

\author[gname='Rin',sname='Miyazaki']{Rin Miyazaki}
\affiliation{Astronomical Institute, Tohoku University, Aoba-ku, Sendai 980-8578, Japan}
\email[show]{rin.miyazaki@astr.tohoku.ac.jp}  

\author[orcid=0000-0002-9053-860X,gname=Masashi, sname='chiba']{Masashi Chiba} 
\affiliation{Astronomical Institute, Tohoku University, Aoba-ku, Sendai 980-8578, Japan}
\email{chiba@astr.tohoku.ac.jp}

\author[orcid=0000-0001-5251-2284, gname=Shunichi,sname=Horigome]{Shunichi Horigome}
\affiliation{Astronomical Institute, Tohoku University, Aoba-ku, Sendai 980-8578, Japan}
\email{shunichi.Horigome@astr.tohoku.ac.jp}

\collaboration{all}{}

\begin{abstract}

We present a spherical Schwarzschild orbit-superposition framework for constraining the dark matter density profiles and stellar orbital structures of the Milky Way dwarf spheroidal galaxies. We model the line-of-sight velocity distribution with a Gaussian mixture whose parameters vary smoothly with projected radius through a mixture density network trained on unbinned stellar velocities. This approach reduces the effect of statistical fluctuations inherent in velocity moments on their kinematic constraints. Their covariance and smoothing-induced correlations are included, and the halo parameters are sampled with Markov Chain Monte Carlo. Mock tests for this model show that the density profiles and slopes near the half-light radius are generally recovered well, although the inferred velocity anisotropy shows some systematic deviations. Applying the model to Sculptor, Draco, and Ursa Minor, we find that the enclosed mass $M_h$ and logarithmic density slope $S_h$ at the projected half-light radius are less sensitive to the adopted two different priors than the asymptotic inner slope $\gamma$, consistent with earlier studies. We further compare the inferred slopes at the three-dimensional half-light radius $r_h$ and $0.5r_h$ for a given stellar mass with FIRE-2 $\Lambda$ cold dark matter ($\Lambda$CDM) and self-interacting dark matter (SIDM) cosmological simulations including baryonic feedback. This comparison avoids introducing halo masses or virial radii that are not directly observable. The inferred slopes are broadly consistent with FIRE-2 $\Lambda$CDM predictions, although current uncertainties and the limited SIDM sample do not yet allow a robust distinction between dark matter models.

\end{abstract}



\section{Introduction}

The $\Lambda$ cold dark matter ($\Lambda$CDM) cosmological model has been remarkably successful in explaining a wide range of cosmological and astrophysical observations on large scales, including the cosmic microwave background (e.g., \citealt{2011ApJS..192...18K}; \citealt{2020A&A...641A...6P}) and the large-scale distribution of galaxies (e.g., \citealt{2006PhRvD..74l3507T}). On smaller scales, however, a number of tensions between $\Lambda$CDM predictions and observations have been discussed, particularly in the regime of dwarf galaxies (see \citealt{2017ARA&A..55..343B}; \citealt{2019A&ARv..27....2S} for reviews). One of the well-known examples is the core-cusp problem: dark-matter-only simulations predict centrally cusped halo density profiles, commonly described by the Navarro--Frenk--White (NFW) profile (\citealt{1997ApJ...490..493N}), whereas some observations of gas-rich dwarf galaxies have suggested the presence of constant-density cores (e.g., \citealt{2015AJ....149..180O}).

The interpretation of the core--cusp problem within the $\Lambda$CDM model has become more complicated when baryonic physics in cosmological simulations is taken into account. Repeated star formation and supernova feedback can transfer energy to the dark matter and produce shallower central density profiles, although the efficiency of this process depends strongly on galaxy mass and star formation history (e.g., \citealt{2014MNRAS.437..415D}; \citealt{2016MNRAS.456.3542T}). Consequently, the inference of a core does not necessarily imply a failure of $\Lambda$CDM. At the same time, alternative dark matter models, such as self-interacting dark matter (SIDM: e.g., \citealt{1992ApJ...398...43C}; \citealt{2000PhRvL..84.3760S}) and fuzzy dark matter (e.g., \citealt{2000PhRvL..85.1158H}; \citealt{2021A&ARv..29....7F}), can also produce cored density profiles. It is therefore difficult to determine whether an observed core reflects baryonic processes or the nature of dark matter. From this perspective, lower-mass systems are particularly valuable because baryonic effects are expected to become less important.

Milky Way dwarf spheroidal galaxies (dSphs), including ultra-faint dwarf galaxies (UFDs), provide important laboratories for probing dark matter. They are largely dark-matter dominated and generally contain little or no gas at the present day (e.g., \citealt{1998ARA&A..36..435M}). UFDs are particularly valuable because their low stellar masses imply weak baryonic effects, but their small stellar populations limit detailed dynamical analyses. Classical dSphs provide a complementary opportunity: although baryonic effects may be more important, their substantially larger kinematic samples allow the dark matter distribution and stellar orbital structure to be constrained in greater detail.

A variety of dynamical methods have been developed to infer the dark matter distribution of dSphs from stellar kinematics (see \citealt{2013NewAR..57...52B}; \citealt{2021MNRAS.501..978R}). Jeans modeling, i.e. by solving the Jeans equations for stellar dynamics (e.g., \citealt{2018MNRAS.481..860R}; \citealt{2020ApJ...904...45H}; \citealt{2026A&A...705A.212B}) is widely used because it is relatively simple and computationally inexpensive. However, the Jeans equations constrain only the velocity moments of the stellar distribution and do not by themselves guarantee that the resulting moment profiles correspond to a physically admissible, non-negative distribution function (DF). In addition, Jeans analyses usually adopt a prescribed functional form for the velocity anisotropy, $\beta(r)=1 - (\sigma_\theta^2 + \sigma_\phi^2)/2\sigma_r^2$, where ($\sigma_r$, $\sigma_\theta$, $\sigma_\phi$) are the velocity dispersions in the spherical coordinates of ($r, \theta, \phi$). This simplified assumption for $\beta(r)$ can affect the derivation of inferred dark matter profile. DF modeling (e.g., \citealt{2019MNRAS.482.1525V}; \citealt{Pascale2020}) avoids the first of these issues by parameterizing the DF itself and computing the observables by integrating over phase-space coordinates. However, the assumed functional form of the DF can still restrict the range of allowed orbital structures and introduce model dependence.

Schwarzschild modeling (\citealt{1979ApJ...232..236S}) provides a less restrictive alternative. In this approach, the observables are reproduced by superposing stellar orbits in a given gravitational potential, while the weights of the orbits are determined without imposing an explicit parametric form on the DF. The orbital weights can then be used to recover the velocity anisotropy and other properties of the orbital structure of stars. This flexibility comes at a high computational cost because a large number of orbits must be integrated for each trial potential. For dSphs, previous Schwarzschild studies have often adopted simple $\chi^2$ grid searches over the dark matter parameters for the sake of simplicity (e.g., \citealt{2013MNRAS.433.3173B}; \citealt{2013ApJ...763...91J}).

In this work, we develop a Schwarzschild modeling framework for constraining the internal dark matter structure of Milky Way dSphs. This orbit-based model incorporates high-order velocity moments, but these are generally sensitive to statistical fluctuations when estimated from sparsely sampled spatial bins. We thus construct smooth kinematic constraints from the individual stellar velocities. Specifically, we model the line-of-sight velocity distribution (LOSVD) as a three-component Gaussian mixture whose parameters are predicted as smooth functions of projected radius by a mixture density network (MDN: \citealt{Bishop1994MixtureDN}) trained on the unbinned stellar data. With this procedure, we derive smooth second- and fourth-order velocity-moment profiles from the resulting LOSVD and use them as kinematic targets in the Schwarzschild modeling, accounting for both their covariance and the correlations introduced by the smoothing process. We further combine the Schwarzschild calculation with Markov Chain Monte Carlo (MCMC) sampling of the dark matter halo parameters, enabling posterior exploration beyond a discrete grid search. Our Schwarzschild model is then tested with mock data, and its validity is confirmed except for remaining systematics for some model parameters.

Based on this framework, we constrain the dark matter density profiles and stellar orbital structures of Sculptor, Draco, and Ursa Minor, and examine how the inferred halo properties depend on the adopted prior. Previous Schwarzschild analyses have shown that the dark matter density slope can be constrained nearly independently of the assumed halo model at a finite radius around $r_{-3}$, where the logarithmic slope of the stellar density profile is $-3$ \citep{2013A&A...558A..35B}. In this context, we compare the asymptotic inner density slope $\gamma$ with the logarithmic density slope evaluated at the projected half-light radius ($R_h$), $S_h \equiv -d\ln\rho/d\ln r|_{R_h}$, and find that $S_h$ is substantially less sensitive to the two prior choices considered here than $\gamma$. We further compare the inferred density slopes at the three-dimensional half-light radius, $r_h$, and at $0.5r_h$ with predictions from CDM and SIDM cosmological simulations including baryonic feedback as a function of stellar mass $M_*$. By expressing this comparison in terms of $M_*$ and radius $r_h$ tied to the measurable stellar distribution, we avoid introducing quantities of dark halos such as the total halo mass $M_{\rm halo}$ and the virial radius $R_{\rm vir}$ that are not directly available from the stellar data. The inferred slopes are broadly consistent with the range predicted by the CDM simulations, although the current uncertainties and the limited set of available SIDM simulations do not yet allow a robust distinction between dark matter models.

The paper is organized as follows. Section~2 describes our density models, Schwarzschild modeling framework combined with an MDN-based treatment of the LOSVD. In Section~3, we first summarize the validation of our method using mock data, and then compare our result for Sculptor with the previous Schwarzschild analysis of \citet{2013MNRAS.433.3173B} (hereafter referred to as B13). We subsequently apply the method to Sculptor, Draco, and Ursa Minor using recent kinematic data sets. Section~4 discusses the prior sensitivity of the inferred halo parameters, the role of higher-order velocity information and orbital anisotropy, and the comparison of the inferred finite-radius density slopes with cosmological simulations. We also discuss the main limitations and future prospects of the present analysis. Finally, we summarize our conclusions in Section~5.

\section{Method}
\label{method}
Our Schwarzschild method follows the basic recipe developed by many previous studies (e.g., \citealt{1997ApJ...488..702R}; B13; \citealt{2020ApJ...889...39V}). We make further improvements to run MCMC within the Schwarzschild modeling framework. 

The core of Schwarzschild's orbit-superposition method is to construct a representative set of physically allowed stellar trajectories (an orbit library) within a given gravitational potential. In a steady-state stellar system, the observed density and kinematic properties are reproduced by assigning appropriate positive weights to individual orbits in this library. The basic flow of Schwarzschild modeling proceeds in four distinct phases: (1) set the density profile and calculate the corresponding potential, (2) Given that potential, integrate the orbit using a given set of integrals of motion, such as energy and angular momentum, (3) project the integrated coordinates onto the observational space, and (4) execute a fitting to obtain posterior distribution of parameters.

\subsection{Models for Dark Matter and Stellar Density Profiles}
To calculate the gravitational potential that governs orbits, we model the corresponding density profile of the dark matter halo.
In this study, we adopt a spherical double power-law profile (\citealt{2008gady.book.....B}, see their equation(2.64)). The density distribution $\rho(r)$ of this model is defined using four parameters as follows:
\begin{equation}
    \label{DM density}
    \rho(r) = \rho_0 \left(\frac{r}{r_0}\right)^{-\gamma} \left(1 + \frac{r}{r_0}\right)^{\gamma - \kappa}
\end{equation}
where $\rho_0$ is the characteristic scale density, and $r_0$ is the scale radius. The parameter $\gamma$ controls the inner density slope near the galactic center, while $\kappa$ determines the outer slope at larger radii. This single mathematical framework encompasses a wide range of density profiles, including the NFW profile ($\kappa = 3, \gamma = 1$) predicted by the $\Lambda$CDM model, and cored profiles with a constant central density ($\gamma = 0$). Since our target galaxies are dark matter dominant dSphs, we calculate the gravitational potential $\Phi(r)$ from only this dark matter density profile.

To reproduce the observed distribution of stars using the calculated orbits, we also model the surface density of the stars. For the stellar density profile, we adopt a  generalized $\alpha\beta\gamma$-profile (\citealt{1996MNRAS.278..488Z}) for the surface density distribution $\Sigma(R)$ as a function of the projected radius $R$:
\begin{equation}
\label{stellar_surface_density}
    \Sigma(R) = \Sigma_0 \left( \frac{R}{R_*} \right)^{-\gamma_*} \left[ 1 + \left( \frac{R}{R_*} \right)^{\alpha_*} \right]^{\frac{\gamma_* - \beta_*}{\alpha_*}}
\end{equation}
where $\Sigma_0$ is a normalization constant, $R_*$ is the scale radius, and $\alpha_*$, $\beta_*$, and $\gamma_*$ control the shape of the profile. We determine these parameters by fitting equation~(\ref{stellar_surface_density}) to the observed projected stellar positions; details of the fitting procedure are given in Appendix~\ref{stellar_fitting}. The projected half-light radius $R_h$ is then calculated from the cumulative projected stellar mass of the fitted profile as the radius enclosing half of the total stellar mass. We normalize $\Sigma(R)$ such that its integral over the adopted radial range is unity.

Assuming spherical symmetry, the three-dimensional stellar density profile $\rho_*(r)$ can be reconstructed from the projected surface density $\Sigma(R)$ via the Abel inversion (\citealt{2008gady.book.....B}):
\begin{equation}
    \label{rho_star}
    \rho_*(r) = -\frac{1}{\pi} \int_r^\infty \frac{d\Sigma(R)}{dR} \frac{1}{\sqrt{R^2 - r^2}} dR
\end{equation}

\subsection{Orbit Library Generation}
\label{Orbit Library Generation}

\subsubsection{Sampling in the ($E, L$) Space}

Under the assumption of spherical symmetry, Jeans' theorem (\citealt{1916MNRAS..76..552J}; \citealt{2008gady.book.....B}) states that the distribution function of a steady-state stellar system can be written as a function of the integrals of motion, $f(\bm{x},\bm{v}) = f(E,L)$, where $E$ and $L$ are the orbital energy and angular momentum, respectively. Thus, individual orbits in a spherical potential can be characterized by a pair $(E,L)$. We construct the orbit library by discretizing this $(E,L)$ space into a grid,
\begin{equation}
    (E_i,L_{ij}),
    \qquad
    i=1,\ldots,N_E,\quad
    j=1,\ldots,N_L,
\end{equation}
where $N_E=50$ and $N_L=20$ in our fiducial setup. The sampling of this grid is designed to cover the relevant phase space while providing sufficient resolution in the regions where the tracer stars are distributed.

We parameterize the orbital energy by the radius of a circular orbit, $r_c(E)$, in the given potential. We sample $r_c$ over the range $0.1R_h \leq r_c \leq 20R_h$, where $R_h$ is the half-light radius of the stellar distribution.
Rather than using a simple linear or logarithmic spacing, we adopt a hybrid grid that combines logarithmic radial spacing with the cumulative mass distribution of the tracer stars. This provides sufficient resolution over the full radial range while allocating more sampling points to regions containing a larger fraction of the tracer population.

For the hybrid grid, we define a dimensionless radial coordinate
$u(r)\in[0,1]$ as
\begin{equation}
    u(r)
    =
    w_{\log}u_{\log}(r)
    +
    (1-w_{\log})u_{\rm mass}(r),
\end{equation}
where
\begin{equation}
    u_{\log}(r)
    =
    \frac{\ln(r/r_{\min})}
    {\ln(r_{\max}/r_{\min})},
\end{equation}
and
\begin{equation}
    u_{\rm mass}(r)
    =
    \frac{\int_0^r \rho_*(s)s^2\,ds}
    {\int_0^{r_{\max}}\rho_*(s)s^2\,ds}.
\end{equation}
Here, $\rho_*(r)$ is the three-dimensional stellar density profile calculated from equation~\eqref{rho_star}, and $r_{\min}=0.1R_h$ and $r_{\max}=20R_h$. We set $w_{\log}=0.5$, giving equal weight to the logarithmic and mass-weighted components.

In practice, $u(r)$ is evaluated on a fine radial grid and the interval $[0,1]$ is divided into $N_E$ equally spaced points. The corresponding values of $r_c$ are obtained by inverse interpolation. This procedure defines the energy grid $\{r_{c,i}\}_{i=1}^{N_E}$ and places more sampling points in regions where the tracer density is higher than would be obtained from logarithmic spacing alone.

For each $r_{c,i}$, the corresponding circular-orbit energy is $E_i, (i = 1, \ldots, N_E)$ and the maximum angular momentum $L_{\rm max}$ for a given energy $E_i$ is that of the circular orbit with this energy. We then define the dimensionless angular momentum $\eta \equiv L / L_{\rm max}(E_i)$.
For each energy, we divide the interval $\eta\in[0.01,0.99]$ into $N_L$ equally spaced cells and use the cell
centers as the sampling points. In the present analysis, the sampling is linear in $\eta$ and the angular momentum grid is given as $L_{ij} = \eta_j L_{\rm max}(E_i),  (j =1, \ldots, N_L)$

\subsubsection{Orbit Integration and Projection to Observational Space}
For each orbit specified by $(E_i,L_{ij})$, we integrate the equations of motion in the spherically symmetric potential. The integration is performed using the adaptive Runge--Kutta--Fehlberg method (RKF45) implemented in the GNU Scientific Library (GSL: \citealt{1538674}). Each integration is initialized at the apocenter $r_a$ and carried out over an orbital period $T_{\text{orb}} = 2\pi r_{\text{a}}/v_{\text{circ}}(r_{\text{a}})$, where $v_{\text{circ}}(r_{\text{a}})$ is the circular velocity evaluated at the $r_a$.

The three-dimensional orbits must be projected onto the plane of the sky to compare with the observed kinematic data (\citealt{1997ApJ...488..702R}). Owing to the assumption of spherical symmetry, the orientation of an orbit relative to the line of sight can be averaged over an isotropic distribution of viewing directions. In practice, for each orbit and each time step, we randomly sample $n_{\mathrm{dir}}=25$ line-of-sight directions. The polar angle $\theta$ between the line of sight and the radial direction is sampled isotropically by drawing $\cos\theta$ uniformly from $[-1,1]$, while the azimuthal angle $\xi$ specifying the orientation of the tangential velocity relative to the line of sight is drawn uniformly from $[0,2\pi)$.

For each orbit, the equations of motion are integrated over the time interval $T_{\mathrm{orb}}$. We sample this interval at $n_{\mathrm{out}}=1200$ equally spaced intervals, resulting in $n_{\mathrm{out}}+1$ time samples including both endpoints. At each time sample, the projected radius and line-of-sight velocity are calculated as $R_{\mathrm{proj}}(t) = r(t)\sin\theta$ and $v_{\mathrm{los}}(t) =v_r(t)\cos\theta - v_t(t)\cos\xi\sin\theta$,
where $v_t=L/r(t)$ is the tangential velocity. The resulting values of $R_{\mathrm{proj}}$ and $v_{\mathrm{los}}$ are then assigned to the corresponding projected radial bins.

For an orbit labelled by $(i,j)$, we accumulate the number of samples $B_{ij,k}$, falling in the $k$-th projected radial bin, $R_k \leq R_{\mathrm{proj}} < R_{k+1}$, together with the corresponding second- and fourth-order velocity moments, denoted as $V_{ij,k}^{(2)}$ and $V_{ij,k}^{(4)}$ respectively. In the continuum limit, these quantities can be expressed as
\begin{equation}
\begin{aligned}
    B_{ij,k} &=
    \frac{1}{T_{\mathrm{orb}}}
    \int_0^{T_{\mathrm{orb}}}
    \mathcal{I}_k(R_{\mathrm{proj}}(t))\,dt,
    \\
    V^{(2)}_{ij,k} &=
    \frac{1}{T_{\mathrm{orb}}}
    \int_0^{T_{\mathrm{orb}}}
    v_{\mathrm{los}}^2(t)
    \mathcal{I}_k(R_{\mathrm{proj}}(t))\,dt,
    \\
    V^{(4)}_{ij,k} &=
    \frac{1}{T_{\mathrm{orb}}}
    \int_0^{T_{\mathrm{orb}}}
    v_{\mathrm{los}}^4(t)
    \mathcal{I}_k(R_{\mathrm{proj}}(t))\,dt,
\end{aligned}  
\end{equation}
where $\mathcal{I}_k$ is the indicator function, which is unity when $R_{\mathrm{proj}}$ falls within the $k$-th radial bin and zero otherwise. In the numerical implementation, these integrals are approximated by the discrete sums over the $n_{\mathrm{out}}+1$ time samples and $n_{\mathrm{dir}}$ randomly sampled viewing directions.

The resulting orbital contributions are used to construct the orbit superposition matrices for the subsequent quadratic programming (QP) optimization. The orbital weights are then determined by combining these contributions to reproduce the observed surface-density and line-of-sight velocity moments.

\subsection{LOSVD Modeling via Mixture Density Network}
\label{MDN-method}

A critical challenge in Schwarzschild modeling is the treatment of statistical fluctuations in the estimated kinematic moments. Schwarzschild models combine a large number of orbits and therefore have substantial flexibility in reproducing the observed velocity moments. When these moments are estimated independently in radial bins and are used directly as fitting targets, fluctuations arising from the finite number of stars in each bin can be absorbed by the model rather than reflecting the underlying kinematic structure. We therefore construct smooth velocity-moment profiles from the individual stellar measurements before performing the Schwarzschild modeling.

This issue is particularly important for the fourth-order velocity moment, $\mu_4$, which is sensitive to individual high-velocity stars and can therefore become noisy when the number of stars is limited. To obtain stable estimates of both the second- and fourth-order moments while retaining information on the shape of the LOSVD, we model the continuous LOSVD as a function of projected radius using a Mixture Density Network (MDN).

\subsubsection{Gaussian Mixture Model}

We model the conditional LOSVD, $p(v|R)$, using a Gaussian Mixture Model (GMM), in which the velocity distribution is represented as a weighted sum of Gaussian components. A GMM provides a flexible representation of non-Gaussian velocity distributions while retaining an analytically tractable form for the velocity moments.

In this work, we impose symmetry of the LOSVD about the systemic velocity and use three Gaussian components: one central component and two symmetric components with equal weights and opposite velocity offsets. The LOSVD is therefore written as
\begin{equation}
    \label{GMM}
  \footnotesize
\begin{aligned}
 p(v | R) &=
 \pi_1 \mathcal{N}(v | 0, \sigma_1^2 + e^2) \\
 &+
 \frac{1-\pi_1}{2}
 \left[
 \mathcal{N}(v | v_0, \sigma_2^2 + e^2)
 +
 \mathcal{N}(v | -v_0, \sigma_2^2 + e^2)
 \right],
\end{aligned}
\end{equation}
where $\mathcal{N}(v \mid \mu, \sigma^2)$ denotes a one-dimensional Gaussian distribution with mean $\mu$ and variance $\sigma^2$, $\pi_1$ is the mixing weight of the central component, $\sigma_1$ and $\sigma_2$ are the intrinsic dispersion scales, $v_0$ is the velocity offset of the two symmetric components, and $e$ is the measurement uncertainty of individual stellar velocities. The observational uncertainty is incorporated by adding $e^2$ to the
intrinsic variance of each Gaussian component.

The symmetric parameterization guarantees a zero mean velocity while allowing the LOSVD to deviate from a single Gaussian. In particular, the relative weights and widths of the components, together with the velocity offset $v_0$, allow the model to represent velocity distributions with different degrees of kurtosis.

\subsubsection{Neural Network Architecture}
In the MDN approach, the GMM parameters ($\pi_1, \sigma_1, v_0, \sigma_2$) are predicted as continuous functions of a projected radius $R$ by a neural network. The network maps standardized spatial features derived from individual stellar positions to the parameters describing the local LOSVD.

To ensure stable optimization against the strong degeneracy between the overall velocity scale and the distribution shape, we adopt an orthogonalized parameterization where the network predicts the velocity dispersion $\sigma_{\mathrm{tot}}$ independently from dimensionless shape parameters. Furthermore, to accurately capture the non-Gaussian velocity tails without biasing the global velocity scale, the network is trained using a tail-weighted loss function incorporating a stop-gradient operator. We refer the reader to Appendix \ref{MDN_detail} for complete details on the feature standardization and mathematical formulation of our approach. 

\subsubsection{Ensemble Learning and Kinematic Targets}
To reduce the sensitivity of the inferred LOSVD to statistical fluctuations and neural-network initialization, we train an ensemble of $M=128$ independent MDNs. The first model is trained on the original stellar sample, while the remaining models are trained on bootstrap resamples of the data with independent random initializations (see Appendix \ref{MDN_detail} for detail).

For each trained MDN, the second- and fourth-order velocity moments are calculated analytically from the GMM parameters. Because the adopted GMM is symmetric about zero, the second and fourth moments are
\begin{equation}
  \label{GMM_moment}
  \small
\begin{aligned}
  \mu_{2, \rm GMM}(R) &= \pi_1 \sigma_1^2 + (1 - \pi_1)(\sigma_2^2 + v_0^2) \\
  \mu_{4, \rm GMM}(R) &= 3\pi_1 \sigma_1^4 + (1 - \pi_1)(3\sigma_2^4 + 6\sigma_2^2 v_0^2 + v_0^4)
\end{aligned}
\end{equation}
The corresponding velocity dispersion and kurtosis are given by
$\sigma(R)=\sqrt{\mu_{2, \rm GMM}}$ and $\kappa(R)=\mu_{4, \rm GMM}(R)/\mu_{2, \rm GMM}(R)^2$.
Specifically, we use the median velocity dispersion ($\sigma_{\text{med}}$) and median kurtosis ($\kappa_{\text{med}}$) of the ensemble to define the target second- and fourth-order moments at each radial position,
\begin{align}
  \mu_{2}(R)&=\sigma_{\rm med}^2(R)\\
  \mu_{4}(R)&=\kappa_{\rm med}(R)\mu_{2}^2(R).
\end{align}
The resulting target vector is
\begin{equation}
\small
\boldsymbol{y}=
\left(
\mu_2(R_1),\ldots,\mu_2(R_{N_R}),
\mu_4(R_1),\ldots,\mu_4(R_{N_R})
\right)^{T}.
\end{equation}

These resulting smooth moment profiles are used as the kinematic targets in the Schwarzschild modeling described in Section~\ref{section_objective_function}. 

\subsection{Orbit Weight Optimization via Quadratic Programming}
\label{subsec:qp}

Using the constructed orbit library (matrices $\mathbf{B}$, $\mathbf{V}^{(2)}$, and $\mathbf{V}^{(4)}$), alongside the smooth target velocity moments via the MDN ensemble, we determine the non-negative orbital weights $\boldsymbol{w} = (w_1, \dots, w_{N_{\text{orbits}}})^T \in \mathbb{R}_{\ge 0}^{N_{\text{orbits}}}$, where $N_{\text{orbits}} = N_E \times N_L$ is the total number of orbits. Here, each component $w_{ij}$ represents the weight of the orbit on the $(E_i, L_{ij})$ grid. The optimization task is formulated to maximize the reproduction of the observed stellar kinematics and spatial density distribution while satisfying physical boundary conditions. To solve this high-dimensional, constrained optimization problem both rapidly and strictly, we employ Quadratic Programming (QP) using the commercial-grade solver Gurobi \citep{gurobi}.

\subsubsection{From Model to Observables: Linearization of Moments}

Following the general framework described in B13, the normalized projected stellar mass $\mu_{0,k}$ in the $k$-th radial bin, alongside the corresponding model predictions for the second- and fourth-order velocity moments ($\hat{\mu}_{2,k}$ and $\hat{\mu}_{4,k}$), can be formally represented as superpositions of the orbital weights:
\begin{align}
    \hat{\mu}_{0,k} &= \frac{\Delta m_{*,k}}{M_*} = \sum_{i,j}^{N_{\text{orbits}}} w_{ij} B_{ij,k}\\
    \hat{\mu}_{2,k} &= \frac{1}{\hat{\mu}_{0,k}} \sum_{i,j}^{N_{\text{orbits}}} w_{ij} V^{(2)}_{ij,k}\\
    \hat{\mu}_{4,k} &= \frac{1}{\hat{\mu}_{0,k}} \sum_{i,j}^{N_{\text{orbits}}} w_{ij} V^{(4)}_{ij,k}
\end{align}
where $M_*$ is the total stellar mass and $\Delta m_{*,k}$ is the stellar mass in the $k$-th bin. Since the velocity moments are inherently non-linear with respect to the orbital weights $\boldsymbol{w}$ (appearing in both the numerator and denominator), standard QP solvers cannot be directly applied. To resolve this non-linearity and preserve the convex quadratic nature of the optimization, we approximate the denominators using the analytical projected surface density profile $\Sigma_{\text{fit}}(R)$ obtained from our robust fitting procedure (see Appendix \ref{stellar_fitting}):
\begin{align}
    \label{mu2}
    \hat{\mu}_{2,k} &\simeq \frac{1}{\int_{R_k}^{R_{k+1}} 2\pi R \, \Sigma_{\text{fit}}(R) \, dR} \sum_{i,j}^{N_{\text{orbits}}} w_{ij} V^{(2)}_{ij,k}\\
    \label{mu4}
    \hat{\mu}_{4,k} &\simeq \frac{1}{\int_{R_k}^{R_{k+1}} 2\pi R \, \Sigma_{\text{fit}}(R) \, dR} \sum_{i,j}^{N_{\text{orbits}}} w_{ij} V^{(4)}_{ij,k}
\end{align}
This approximation is mathematically justified, as the physical orbital weights are strictly constrained to self-consistently reproduce the target stellar density profile within a tight tolerance, as shown in equation \eqref{constraints}.

\subsubsection{Objective Function}
\label{section_objective_function}

To determine the orbit weights $\boldsymbol{w}$, we minimize the following objective function:
\begin{equation}
  \label{objective_function}
    \mathcal{J}(\boldsymbol{w}) = \mathcal{J}_{\text{kin}}(\boldsymbol{w}) + \mathcal{R}(\boldsymbol{w}) + \lambda_s \sum_{k'=1}^{N_\text{bin}}s_{k'}^2,
\end{equation}
which consists of the kinematic mismatch term, a quadratic regularization term, and a penalty on the slack variables. The kinematic term is defined as
\begin{equation}
    \mathcal{J}_{\text{kin}}(\boldsymbol{w})=\sum_{k=1}^{N_{\text{bin}}} (\boldsymbol{y}_k - \hat{\boldsymbol{y}}_k)^T \mathbf{\Sigma}_k^{-1} (\boldsymbol{y}_k - \hat{\boldsymbol{y}}_k),
\end{equation}
where $\boldsymbol{y}_k = (\mu_{2,k}, \mu_{4,k})^T$ and $\hat{\boldsymbol{y}}_k = (\hat{\mu}_{2,k}, \hat{\mu}_{4,k})^T$ are the target and model-predicted moment vectors for the $k$-th radial bin, respectively. The matrix $\mathbf{\Sigma}_k$ is the $2 \times 2$ local covariance matrix that captures the correlation between the second- and fourth-order moments within each bin.

To suppress overfitting arising from the large number of degrees of freedom in the orbit library and to ensure a smooth, physically realistic phase-space structure, we introduce an $L_2$ regularization term $\mathcal{R}(\boldsymbol{w})$, following \citet{2020ApJ...889...39V}:
\begin{equation}
    \mathcal{R}(\boldsymbol{w}) = \lambda \sum_{i,j}^{N_{\text{orbits}}} \left( \frac{w_{ij}}{w_{ij, \text{prior}}} \right)^2
\end{equation}
where $\lambda$ is the regularization parameter (scaled as $\lambda \approx 5.0 \times N_{\text{orbits}}$ to balance the kinematic constraints), and $w_{ij, \text{prior}}$ represents a uniform prior weight distribution across the orbit library. This quadratic penalty prevents the optimization from concentrating too much mass into a handful of individual trajectories, promoting a smooth and continuous distribution of orbital weights across the accessible phase space.

The third term in equation \eqref{objective_function} corresponds to the near-hard constraint on surface density. For each radial bin, the projected stellar mass predicted by the orbit model, $\hat{\mu}_{0,k'}$, is required to reproduce the target value $\mu_{0,k'} = \int_{R_k}^{R_{k+1}} 2\pi R\,\Sigma_{\mathrm{fit}}(R)\,dR$ to within a fractional tolerance of $\delta=0.035$. To allow for occasional violations of this tolerance, we introduce a non-negative slack variable that represents the excess fractional deviation:
\begin{equation}
s_{k'} =
\max\left[
0,\,
\left|
\frac{\hat{\mu}_{0,k'}-\mu_{0,k'}}
{\mu_{0,k'}}
\right|
-\delta
\right].
\label{constraints}
\end{equation}
Thus, $s_{k'}=0$ when the model reproduces the target surface density within $3.5\%$, while $s_{k'}$ gives the amount by which the fractional deviation exceeds this tolerance. We penalize these excess deviations through the third term of equation \eqref{objective_function} with $\lambda_s=10^9$ in this work. The large penalty strongly suppresses violations of the nominal $3.5\%$ tolerance while allowing the QP problem to remain feasible in exceptional cases.

The orbital weights are additionally constrained to be non-negative and normalized,
\begin{equation}
w_{ij}\geq0,
\qquad
\sum_{i,j}w_{ij}=1.
\end{equation}
These conditions ensure that the orbital superposition corresponds to a non-negative and normalized distribution function.
%

\subsection{Statistical Inference}
\label{sec:statistical_inference}
Motivated by previous studies showing that dark matter properties are more robustly constrained at finite radii \citep{2010MNRAS.406.1220W, 2013A&A...558A..35B} and by recent analyses using the density slope at the half-light radius \citep{2025MNRAS.541.1016G}, we define the dark matter parameter vector as
\begin{equation}
\boldsymbol{\theta}_{\rm DM}
\equiv
\{S_h,\log_{10}M_h,\gamma,\kappa\},
\end{equation}
where $S_h\equiv-d\ln\rho/d\ln r|_{R_h}$ is the logarithmic density slope at the projected stellar half-light radius $R_h$, and $M_h$ is the enclosed dark matter mass within $R_h$. The parameters $\gamma$ and $\kappa$ represent the inner and outer asymptotic density slopes, respectively, as defined in Equation~\eqref{DM density}. Selecting these sampling parameters reduces the autocorrelation time of the MCMC and improves sampling efficiency.

For the double power-law profile adopted here, these slopes are related by
\begin{equation}
S_h=\gamma+(\kappa-\gamma)\frac{R_h}{R_h+r_0},
\end{equation}
so that $S_h$ approaches $\gamma$ only when $R_h\ll r_0$; more generally, the relation between a finite-radius slope $S_h$ and the asymptotic slope $\gamma$ depends on the adopted density-profile parameterization. 

For each set of halo parameters $\boldsymbol{\theta}_{\rm DM}$, the QP optimization described in Section~\ref{subsec:qp} determines the orbital weights that best reproduce the target stellar density and kinematic profiles. The resulting minimum kinematic residual, $\chi^2_{\rm min}(\boldsymbol{\theta}_{\rm DM})=\mathcal{J}_{\text{kin}}(\bm{w})$, is then used to construct the likelihood for the halo parameters. In this section, we describe the likelihood, the treatment of correlations in the MDN-derived kinematic targets, the adopted priors, and the MCMC sampling procedure.

\subsubsection{Likelihood Function and Statistical Corrections}
\label{likelihood_method}
The likelihood is based on the minimum kinematic residual obtained from the QP optimization, $\chi^2_{\rm min}(\boldsymbol{\theta}_{\rm DM})$. The QP objective accounts for the local covariance between the second- and fourth-order velocity moments in each radial bin. However, the MDN preprocessing introduces additional correlations between neighboring radial bins because the velocity-moment profiles are estimated as smooth functions of projected radius. Treating the resulting kinematic targets as independent would therefore overestimate the amount of independent information and lead to overly narrow posterior distributions.

To account for these correlations, we construct a full covariance matrix from the MDN ensemble. Let $\mathbf{C}$ denote the covariance matrix of the combined vector of second- and fourth-order moments over all radial bins. Its elements are given by
\begin{equation}
C_{ab}
=
{\rm cov}(y_a,y_b)
+
\delta_{ab}\,J_a,
\end{equation}
where $a$ and $b$ index both radial bin and moment order, and $J_a$ denotes the additional diagonal variance term described below.

Because the MDN targets are estimated from a finite number of stars in each radial bin, we include a small diagonal variance component to account for the corresponding sampling fluctuations. For a radial bin containing $N_k$ stars, we adopt
\begin{equation}
J_{\mu_2,k}
=
\frac{2}{N_k}\,C_{\mu_2,k,\mu_2,k},
\qquad
J_{\mu_4,k}
=
\frac{8}{N_k}\,C_{\mu_4,k,\mu_4,k},
\end{equation}
where the diagonal variances on the right-hand side are those estimated from the MDN ensemble. These terms provide a shot-noise-like variance floor that scales with the number of stars in each radial bin. They are not intended as an exact analytic Poisson error model, but rather as a conservative treatment of the additional sampling uncertainty associated with finite stellar counts.

For numerical stability, the resulting covariance matrix is subsequently regularized by eigenvalue clipping: eigenvalues smaller than $10^{-6}$ times the largest eigenvalue are replaced by this threshold. The resulting covariance matrix is positive definite and can therefore be used to construct the corresponding correlation matrix,
\begin{equation}
{\rm Corr}_{ab}
=
\frac{C_{ab}}{\sqrt{C_{aa}C_{bb}}}.
\end{equation}

We then characterize the effective number of independent constraints in the MDN-derived kinematic targets using
\begin{equation}
\nu_{\rm eff}
=
\frac{
\left[{\rm tr}(\mathbf{Corr})\right]^2
}{
{\rm tr}(\mathbf{Corr}^2)
}.
\end{equation}
For independent data points, $\nu_{\rm eff}$ approaches the total dimension of the data vector, whereas strong correlations reduce $\nu_{\rm eff}$.

We account for this reduction in independent information by introducing a temperature factor
\begin{equation}
T=
\frac{2N_{\rm bin}}{\nu_{\rm eff}},
\end{equation}
where $2N_{\rm bin}$ is the number of kinematic constraints from the second- and fourth-order moments. The likelihood is then defined as
\begin{equation}
\ln\mathcal{L}(\boldsymbol{\theta}_{\rm DM})
=
-\frac{1}{2}
\frac{\chi^2_{\rm min}(\boldsymbol{\theta}_{\rm DM})}{T}.
\end{equation}
This temperature scaling reduces the effective weight of correlated kinematic constraints and prevents the smooth MDN-derived profiles from artificially providing more independent information than is supported by the stellar sample.

\subsubsection{MCMC Sampling}
\label{subsec:mcmc_sampling}

We consider two choices of prior for $\boldsymbol{\theta}_{\rm DM}$ to assess the sensitivity of the inferred halo properties to the adopted prior. In both cases, the scale radius $r_0$ and characteristic density of the halo $\rho_0$ are restricted to the same range,
\begin{equation}
  \label{range}
\log_{10}r_0
\in
\left[
\log_{10}(0.01r_h),
\log_{10}(50r_h)
\right],
\end{equation}
\begin{equation}
\log_{10}\rho_0 \in [-4,0]
\quad
[{\rm M_\odot\,pc^{-3}}],
\end{equation}
where $r_h$ is the three-dimensional stellar half-light radius. The two priors differ only in how the allowed parameter space within this range is weighted.

In the fiducial case, referred to as \textit{Prior 1}, we impose a uniform prior on the logarithmic density slope at the projected half-light radius, $S_h$. The prior ranges are
\begin{equation}
\gamma < S_h < \kappa,
\end{equation}
\begin{equation}
\log_{10}M_h \in [5,10],
\end{equation}
\begin{equation}
\gamma \in [0,2),
\end{equation}
and
\begin{equation}
\kappa \in [2,10].
\end{equation}
Although the prior is uniform in $S_h$, the scale radius is additionally restricted to the range given above. Consequently, the induced prior weighting in $r_0$ is not uniform.

As an alternative, referred to as \textit{Prior 2}, we replace the uniform prior on $S_h$ with a uniform prior on $\log_{10}r_0$ over the same allowed range of equation \eqref{range}. The priors on $\log_{10}M_h$, $\gamma$, and $\kappa$ are kept identical to those in \textit{Prior 1}. Thus, the two prior choices allow the same range of scale radii but assign different prior weights within that range: \textit{Prior 1} is uniform in $S_h$, whereas \textit{Prior 2} is uniform in $\log_{10}r_0$.

For each sampled set of halo parameters, the orbital weights are optimized by solving the quadratic programming problem described in Section~\ref{subsec:qp}. We then evaluate the likelihood using the minimum $\chi^2$ obtained from the optimized orbital weights. The posterior distribution is sampled using the \texttt{zeus} ensemble slice sampler (\citealt{9646697}). We employ 48 walkers, discard the first 200 steps as burn-in, and retain the subsequent 800 steps for scientific inference. Convergence is assessed using the Gelman--Rubin statistic and the integrated autocorrelation time.

\subsection{Observational Data for Three dSphs}
We analyze three Milky Way dSphs: Sculptor (Scl), Draco (Dra), and Ursa Minor (UMi). For Sculptor, we adopt two different data sets depending on the purpose of our analysis. In Section~\ref{result1}, where we benchmark our results against those of B13, we use the public spectroscopic data set of \citet{2009AJ....137.3100W} (hereafter W09). While B13 utilized a combined data set incorporating both W09 and the data of \citet{2008ApJ...681L..13B}, we restrict our selection to the uniform, publicly available W09 catalog, selecting stars with membership probabilities greater than 0.95. For the stellar density distribution, we adopt a Plummer profile (\citealt{1911MNRAS..71..460P}), which corresponds to equation (2) with $\alpha_\ast=2$, $\beta_\ast=4$, and $\gamma_\ast=0$, i.e., $\Sigma(R)\propto [1+(R/R_\ast)^2]^{-2}$. We set the scale radius to $R_\ast=300$ pc, following the setup of B13.

For the comparison of Sculptor, Draco, and Ursa Minor in Section~\ref{result_recent}, we instead use the most recently available data sets. For Sculptor, we use the spectroscopic catalog of \citet{2023yCat..36750049T} (hereafter T23) and select the stars identified as members in the catalog. For Draco and Ursa Minor, we use the catalogs compiled by \citet{2026A&A...710A.291P} (hereafter P26), available through Zenodo \citep{P26_zenodo_data}. For the stellar density distribution of Sculptor, we adopt a Plummer profile with a scale radius of $R_\ast=280$ pc, following \citet{2018ApJ...860...66M}. For Draco and Ursa Minor, we fit the stellar surface-density distribution using the photometric data provided in the P26 catalogs and adopt the best-fitting parameters of equation~\eqref{stellar_surface_density} (Appendix~\ref{stellar_fitting}).

The data sets, numbers of member stars, stellar density-profile parameters, projected half-light radii $R_h$, and references used in this work are summarized in Table~\ref{setup}. 
\begin{table}
    \centering
    \caption{Summary of Data Sets and Analysis Setups}
    \label{setup}

    \setlength{\tabcolsep}{4pt}
    \begin{tabular}{l c r c c c r r c}
        \tableline
        Galaxy & Data & $N_{\rm m}$ &
        $\alpha_\ast$ & $\beta_\ast$ & $\gamma_\ast$ &
        $R_\ast$ & $R_h$ & $R_{\rm m}$ \\
        \tableline
        Scl$^{(1),(2)}$ & W09 & 1348 & 2    & 4    & 0    & 300 & 300 & 1338 \\
        Scl$^{(3),(4)}$ & T23 & 1604 & 2    & 4    & 0    & 280 & 280 & 1860 \\
        Dra$^{(5)}$     & P26 & 710  & 2.26 & 3.77 & 0.15 & 176 & 182 & 551  \\
        UMi$^{(5)}$     & P26 & 525  & 1.88 & 3.24 & 0.17 & 283 & 386 & 1110 \\
        \tableline
    \end{tabular}

    \tablecomments{
    $N_{\rm m}$ is the number of member stars used in the analysis. The parameters $(\alpha_\ast,\beta_\ast,\gamma_\ast,R_\ast)$ describe the stellar surface-density profile in equation~\eqref{stellar_surface_density}. $R_h$ is the projected half-light radius, and $R_{\rm m}$ is the maximum projected radius of the stars in each data set. The radii $R_\ast$, $R_h$, and $R_{\rm m}$ are given in pc.
    }

    \tablerefs{
    (1) \citet{2009AJ....137.3100W};
    (2) \citet{2013MNRAS.433.3173B};
    (3) \citet{2023yCat..36750049T};
    (4) \citet{2018ApJ...860...66M};
    (5) \citet{2026A&A...710A.291P}.
    }
\end{table}

\section{Results}
\subsection{Validation with Mock Data}
\label{mock_summary}
Before applying our method to the observational data of the Milky Way dSphs, we assess its recovery performance using four mock data sets covering both cuspy and cored dark matter halos and several different orbital structures (Mock 1 - 4 as summarized in Table 2). Each mock catalog contains $1500$ stars and is analyzed using the same modeling framework as adopted for the observed galaxies. The full setup and detailed results of these tests are presented in Appendix~\ref{mock}. Here we briefly summarize the aspects which are most relevant to the interpretation of the observational results.

We find that the MDN-based procedure recovers well the overall radial behavior of the true second- and fourth-order velocity moments for all four mocks, with two representative examples shown in Figure~\ref{MDN}. The second moment is recovered more accurately than the fourth moment, where the latter shows larger deviations owing to its stronger sensitivity to the tails of the LOSVD. For the dynamical inference, the true dark-matter density profiles are broadly recovered within the $2\sigma$ credible intervals for all four mocks (Figure~\ref{Mock_density}). A notable limitation appears for the asymptotic inner slope $\gamma$ in Mock~2, which has a large scale radius ($r_0/R_h=25$ compared with other mocks with $r_0/R_h=4$ or $8$): its posterior peaks clearly away from the true value because the adopted prior weighting disfavors large $r_0$ values (see Appendix~\ref{mock} and Section~\ref{first_discussion} for details). The overall radial behavior of the velocity anisotropy profiles is also recovered within, or close to, the $2\sigma$ credible intervals (Figure~\ref{Mock_beta}). However, the posterior median can deviate systematically from the true profile, with offsets reaching approximately $0.2$--$0.3$ in $\beta$ in the most discrepant radial ranges. The magnitude and radial behavior of these deviations vary among the mocks. Despite these limitations, the logarithmic density slope at the half-light radius, $S_h$, is recovered well in all four cases. In the following applications to the observed galaxies, we keep these recovery properties in mind when interpreting the inferred dark matter density and orbital structure.

\subsection{Application to Sculptor with W09 Data}
\label{result1}
\begin{figure}
    \centering
    \includegraphics[width=\columnwidth]{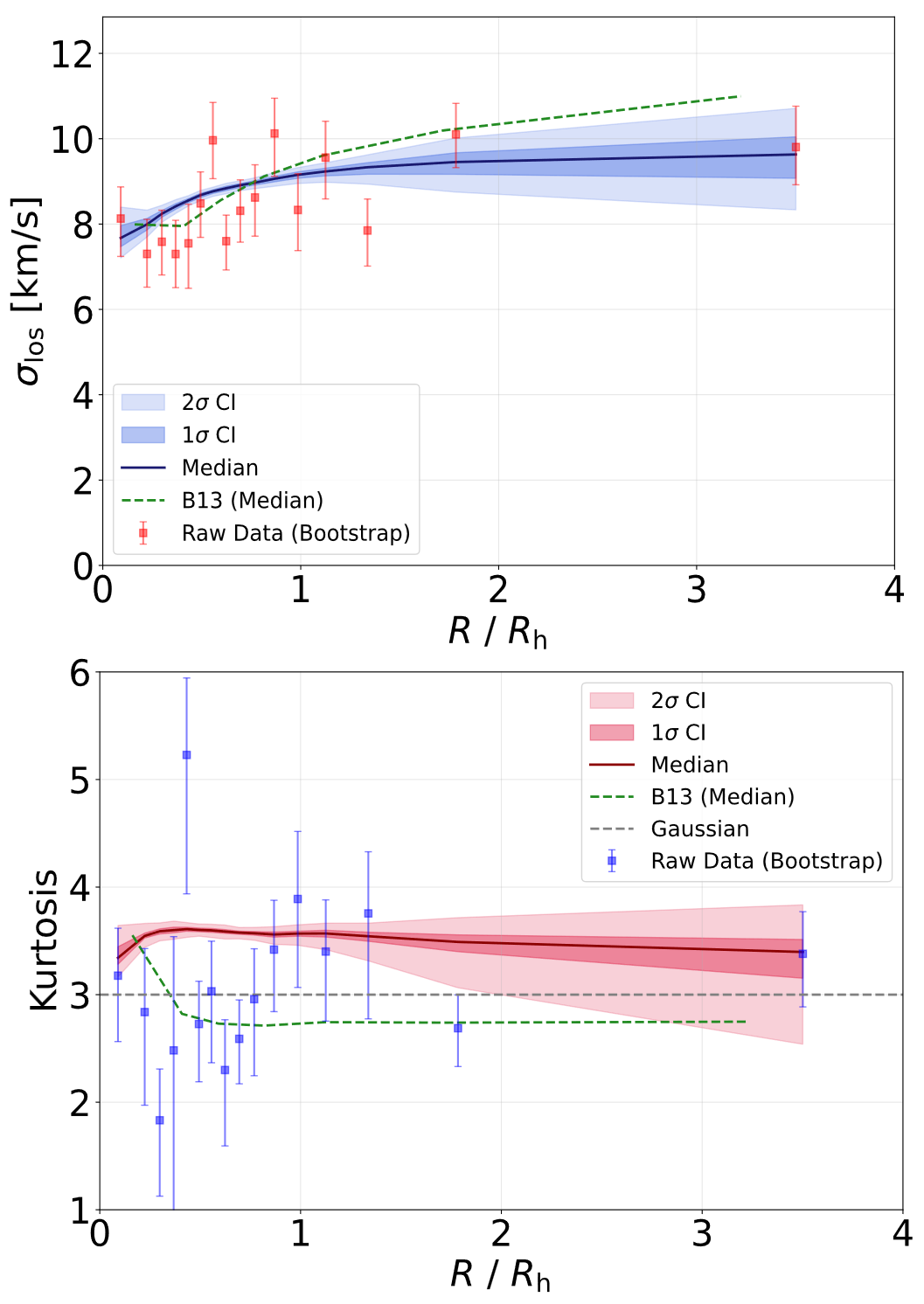}
    \caption{Line-of-sight velocity dispersion (top) and kurtosis (bottom) profiles for Sculptor obtained from the W09 data set. The solid lines show the posterior medians, and the dark and light shaded regions indicate the $1\sigma$ and $2\sigma$ credible intervals, respectively. The green dashed lines show the median profiles from B13. The points with error bars are estimates obtained from the observed data using bootstrap resampling. These points are shown for comparison only and are not used directly in the likelihood; our model uses the smoothed second and fourth velocity moments obtained from the MDN.   
    }
    \label{sigmalos_and_kurtosis}
\end{figure}

We apply our model to the W09 data set for Sculptor for the purpose of comparison with B13's Schwarzschild model. Figure~\ref{sigmalos_and_kurtosis} shows the resulting line-of-sight velocity  dispersion and kurtosis profiles as a function of projected radius. The solid line shows the posterior median, while the dark and light shaded regions  indicate the $1\sigma$ and $2\sigma$ credible intervals, respectively. For comparison, the median profiles obtained by B13 are shown by the green dashed lines. The points with error bars show estimates obtained directly from the data using  bootstrap resampling. Note that these data points are shown only for visualization. In the actual modeling, we use the second and fourth velocity moments obtained  from the smoothed line-of-sight velocity distribution estimated by the MDN,  rather than the binned data points shown in the figure.

\begin{figure*}
    \centering
    \includegraphics[width=\textwidth]{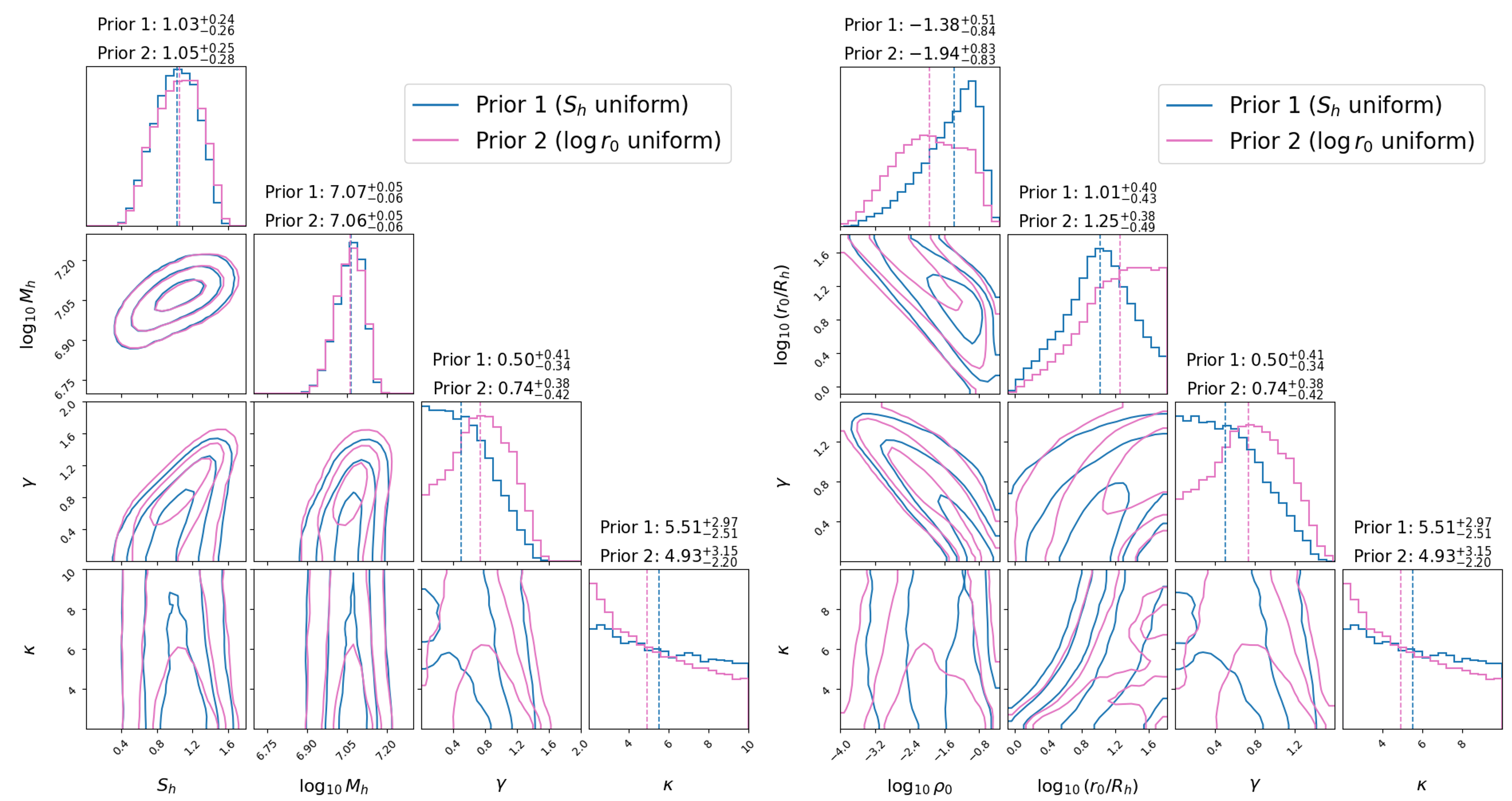}
    \caption{
    Corner plots showing the posterior probability distributions of the sampling parameters (left) and derived dark matter halo parameters (right) for Sculptor using the W09 data set. In each corner plot, the diagonal panels show the one-dimensional marginalized posterior distributions, where vertical dashed lines mark the posterior medians and the values printed above give the median and $1\sigma$ credible intervals. The off-diagonal panels display the two-dimensional joint marginalized posterior distributions with contours representing the credible regions. The blue and magenta lines/contours correspond to the results obtained with Prior 1 (uniform prior on $S_h$) and Prior 2 (uniform prior on $\log_{10} r_0$), respectively.
    }
    \label{pdf_prior_difference}
\end{figure*}

The velocity dispersion profiles are broadly similar between our model and B13,  although some differences are visible at larger radii. In contrast, the kurtosis profiles show a more noticeable difference. The B13 profile appears to be pulled towards specific binned measurements with smaller uncertainties. This is expected in part because their modeling directly uses the observed velocity distribution,  whereas our likelihood is constructed from the smoothed velocity moments. The difference in the kurtosis profile can lead to the different constraints on the velocity anisotropy and dark matter density profile obtained in the two analyses, as discussed below.

Figure~\ref{pdf_prior_difference} shows the posterior distributions of the sampling parameters (left) and the derived parameters of the dark matter density profile (right). The MCMC sampling is performed in the parameter space shown in the left panel, while quantities such as $\rho_0$ and $r_0$ are derived from the sampled parameters. We consider two different choices of prior: a uniform prior on $S_h$ (Prior 1, blue) and a uniform prior on $\log r_0$ (Prior 2, magenta). This comparison allows us to examine how sensitive the inferred parameters are to the choice of prior. The values shown above each one-dimensional posterior are the median and $1\sigma$ credible interval, and the dashed vertical lines indicate the corresponding median values.

The mass within the half-light radius, $M_h$, is well constrained within a $\pm 0.05-0.06\ \mathrm{dex}$ ($\log_{10} M_h = 7.07_{-0.06}^{+0.05}$ and $7.06_{-0.06}^{+0.05}$ for the two priors, respectively), showing minimal sensitivity to the choice of prior considered here. This is consistent with previous results from spherical dynamical modeling, which have shown that the enclosed mass near the half-light radius can be constrained robustly from stellar kinematics (e.g., \citealt{2009ApJ...704.1274W}; \citealt{2010MNRAS.406.1220W}). The posterior of the logarithmic density slope at the half-light radius, $S_h$, is also insensitive to the choice of prior. The median changes from $1.03$ to $1.05$, while the width of the posterior remains similar. This behavior is consistent with the earlier finding of \citet{2013A&A...558A..35B} that the dark matter density slope is nearly model-independent at a finite radius around $r_{-3}$.

In contrast, as shown in the right corner plot, the density parameters ($\rho_0, r_0, \gamma, \kappa$) are less well constrained, with uncertainties comparable to their median values and their posterior distributions show a stronger dependence on the choice of prior. In particular, the peak of the $\gamma$ posterior shifts significantly from around 0.7 to 0 between the two prior cases. This suggests that these parameters are not robustly determined by the line-of-sight velocity data alone, at least within the range of priors considered here. Consequently, due to the limited constraints from the data, these posterior distributions may be driven by the shape of the prior (as discussed in Section \ref{first_discussion}). This behavior is also consistent with the mock tests in Section~\ref{mock_summary}, in which $S_h$ is recovered well whereas $\gamma$ and the detailed inner density profile are less robust.

\begin{figure}
    \centering
    \includegraphics[width=\columnwidth]{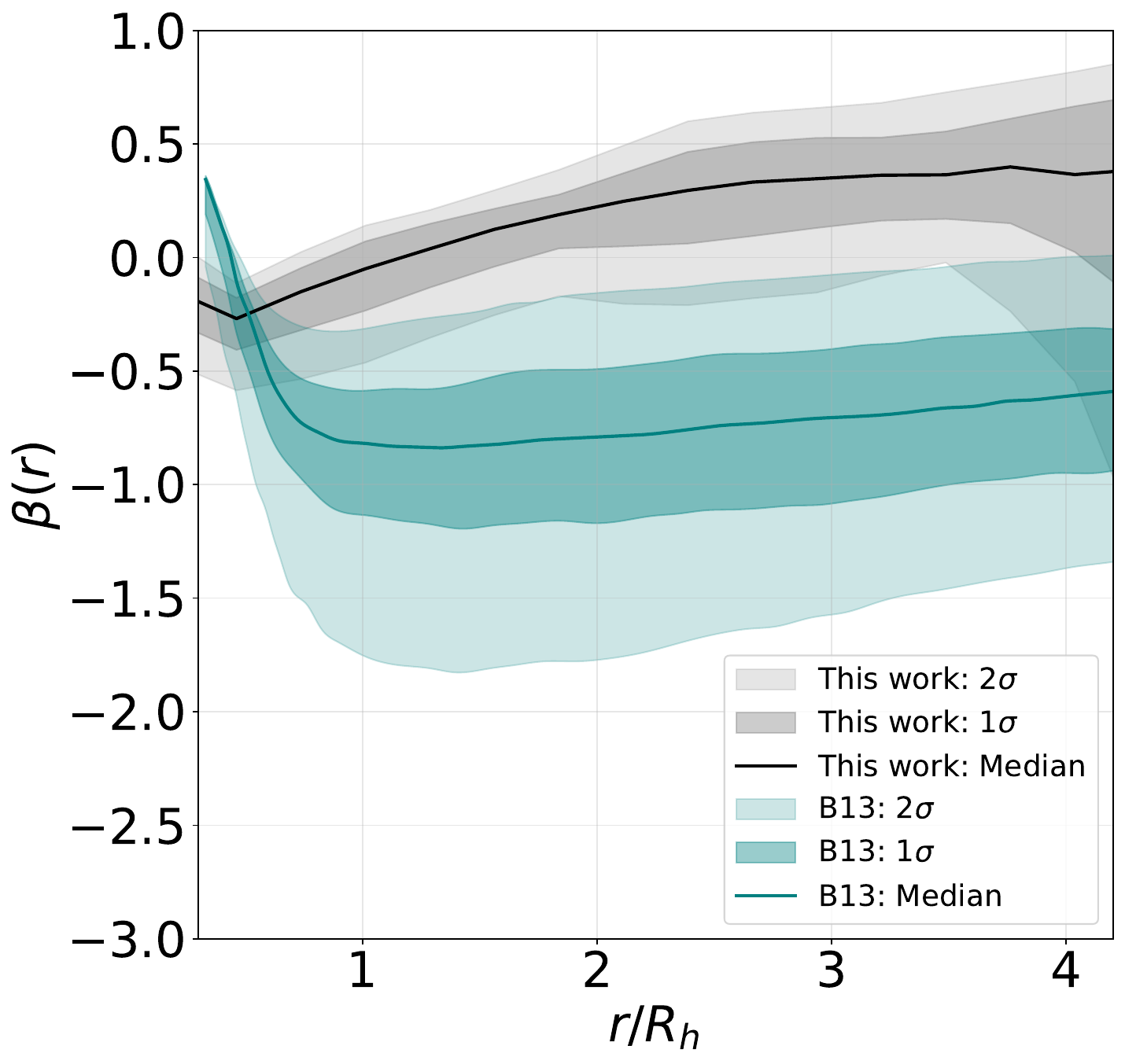}
    \caption{Velocity anisotropy profile for Sculptor obtained from the W09 data set with Prior 1. The black and cyan lines show the posterior medians from this work and B13, respectively. The dark and light shaded regions indicate the $1\sigma$ and $2\sigma$ credible intervals. The radius is normalized by the projected half-light radius $R_h$. Note that the obtained profile with prior2 is almost identical to that obtained with Prior 1.
    }
    \label{beta_W2009}
\end{figure}

Figure~\ref{beta_W2009} shows the inferred velocity anisotropy profile as a function of radius, normalized by the half-light radius $R_h$. The gray and orange regions show the results of this work and B13, respectively. The solid lines indicate the posterior medians, while the dark and light shaded regions correspond to the $1\sigma$ and $2\sigma$ credible intervals.

Our result yields a narrower posterior for the anisotropy profile than that obtained by B13. One possible reason is that our modeling uses smoothed velocity moments estimated with the MDN, rather than fitting directly to the noisier binned estimates of the velocity moments. The smoothing therefore reduces fluctuations in the input kinematic profiles and leads to a narrower posterior. 

We also find that the inferred anisotropy profile is insensitive to the choice of prior. The result obtained with Prior 2, a uniform prior on $\log r_0$, is almost indistinguishable from that obtained with Prior 1, and is therefore not shown in the figure. This indicates that the anisotropy profile is not strongly affected by the choice of prior considered here.

The radial profile of the velocity anisotropy differs substantially from that found by B13. B13 inferred a predominantly tangential orbital structure over the radial range shown, whereas our result is mildly tangential in the central region and becomes increasingly radially biased toward larger radii. The difference between the two median profiles is larger than the typical systematic offsets of $\sim0.2$--$0.3$ in $\beta$ found in our mock tests over much of the radial range. Therefore, this discrepancy is unlikely to be explained solely by the recovery bias identified in the mock analysis and may instead reflect methodological differences between the two analyses. In particular, the different kurtosis profiles shown in Figure~\ref{sigmalos_and_kurtosis} may contribute to the difference, since the fourth-order velocity information affects the inferred orbital structure. This comparison therefore illustrates the sensitivity of the recovered anisotropy to the treatment of higher-order kinematic information, as discussed further in Section~\ref{second_discussion}.

At large radii, the uncertainty in our anisotropy profile becomes larger. This behavior is expected because the outermost radial bins contain little kinematic information. In contrast, the uncertainty reported by B13 becomes narrower toward larger radii, even in a region where the observational data are sparse. The origin of this difference likely arises from a methodological limitation: B13 enforces a heavy dithering scheme that groups $8 \times 8 = 64$ orbits to share the same orbit weight, reducing the distribution function grid from $N_E \times N_L = 160\times64=10240$ to only $20\times8=160$ independent degrees of freedom. This treatment restricts orbital flexibility and might narrow the credible interval in the outer region.

\begin{figure}
    \centering
    \includegraphics[width=\columnwidth]{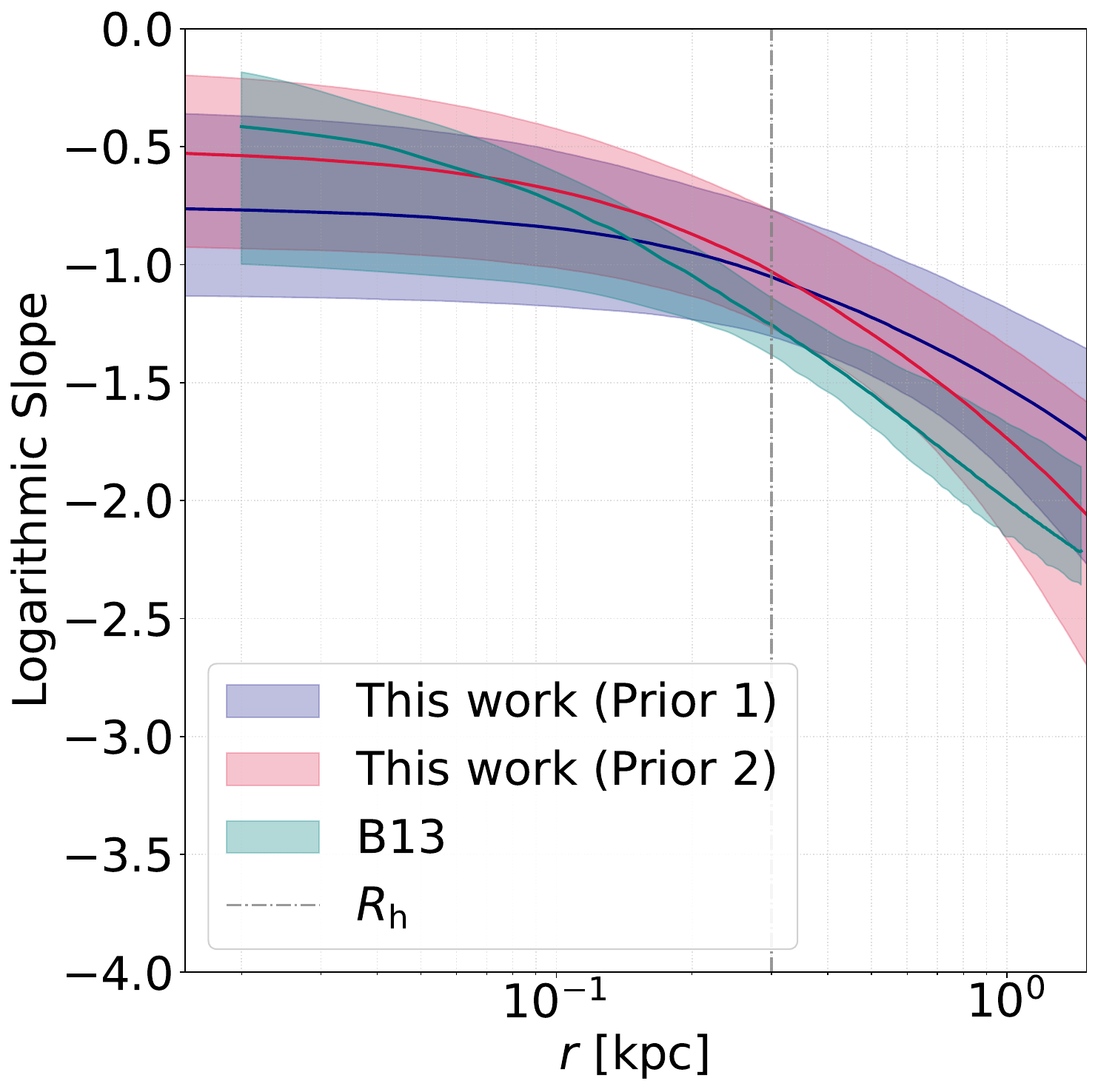}
    \caption{Logarithmic slope of the dark matter density profile as a function of radius for Sculptor using the W09 data set. The purple and red curves show the results obtained with Prior 1 and Prior 2, respectively, while the cyan curve shows the result from B13. The solid lines indicate the posterior medians and the shaded regions show the $1\sigma$ credible intervals. The vertical dash-dotted line indicates the projected half-light radius.
    }
    \label{logslope_W2009}
\end{figure}

Figure~\ref{logslope_W2009} shows the logarithmic slope of the dark matter
density profile, ${d\ln\rho}/{d\ln r}$, as a function of radius. The purple and red curves show the results obtained with Prior 1 and Prior 2, respectively, while the cyan curve shows the result from B13. The solid lines indicate the posterior medians, and the shaded regions show the $1\sigma$ credible intervals.
In the innermost region, the two priors yield different posterior medians for the inferred slope while maintaining comparable uncertainty widths. Around the half-light radius, in contrast, the two priors produce nearly identical posterior distributions, with the uncertainty reaching a minimum in this region. This behavior is consistent with the result shown in Figure~\ref{pdf_prior_difference}, where $S_h$, the logarithmic density slope at the half-light radius, is insensitive to the choice of prior. The B13 profile in the central region is consistent with our result using Prior 2 (a uniform prior on $\log r_0$), with their credible intervals substantially overlapping. 

At the half-light radius, the B13 profile is somewhat steeper than our results. This difference may be related to the different velocity anisotropy profiles inferred in the two analyses as shown in Figure~\ref{beta_W2009}. B13's model  shows a more tangentially biased orbital structure than our model, whereby the degeneracy between velocity anisotropy and the dark matter density slope leads to different inferred density profiles along the corresponding degeneracy direction (as discussed in Section \ref{second_discussion}). 

At large radii, the uncertainty in our inferred slope becomes increasingly large. In contrast, the uncertainty in B13's model remains relatively narrow toward the outer region. B13 fixes the outer logarithmic slope to $\kappa=3$, which naturally leads to a substantially stronger constraint on the density profile at large radii. This difference in the treatment of the outer slope also affects the inferred slope around the half-light radius.

However, these interpretations should be treated with caution, as the two analyses adopt numerous methodological differences—ranging from fundamental assumptions and the treatment of higher-order velocity moments discussed above to finer details like parameterization.

\subsection{Application to Sculptor, Draco, Ursa Minor with Recent Data Sets}
\label{result_recent}
We next apply our spherical Schwarzschild model to Sculptor, Draco, and Ursa Minor using recent kinematic data sets. In this section, we compare our results with previous spherical dynamical models, rather than restricting the comparison to previous Schwarzschild models. This comparison includes the Jeans analysis of \citet{2019MNRAS.484.1401R} (hereafter R19), the DF modeling of \citet{Pascale2020} (hereafter P20), and the simulation-based inference approach of \citet{2026arXiv260626218N} (hereafter N26) for Draco. For all three galaxies, we adopt a uniform prior on $S_h$ (Prior 1). It is important to note that each of these previous works does not use identical kinematic data sets. Differences in sample selection, spatial coverage, and measurement uncertainties may therefore contribute to the differences among the inferred profiles. In the following, however, we focus primarily on how differences in the adopted dynamical models and inference methods may affect the results, while keeping in mind the recovery performance found in the mock tests.
\begin{figure*}
    \centering
    \includegraphics[width=\textwidth]{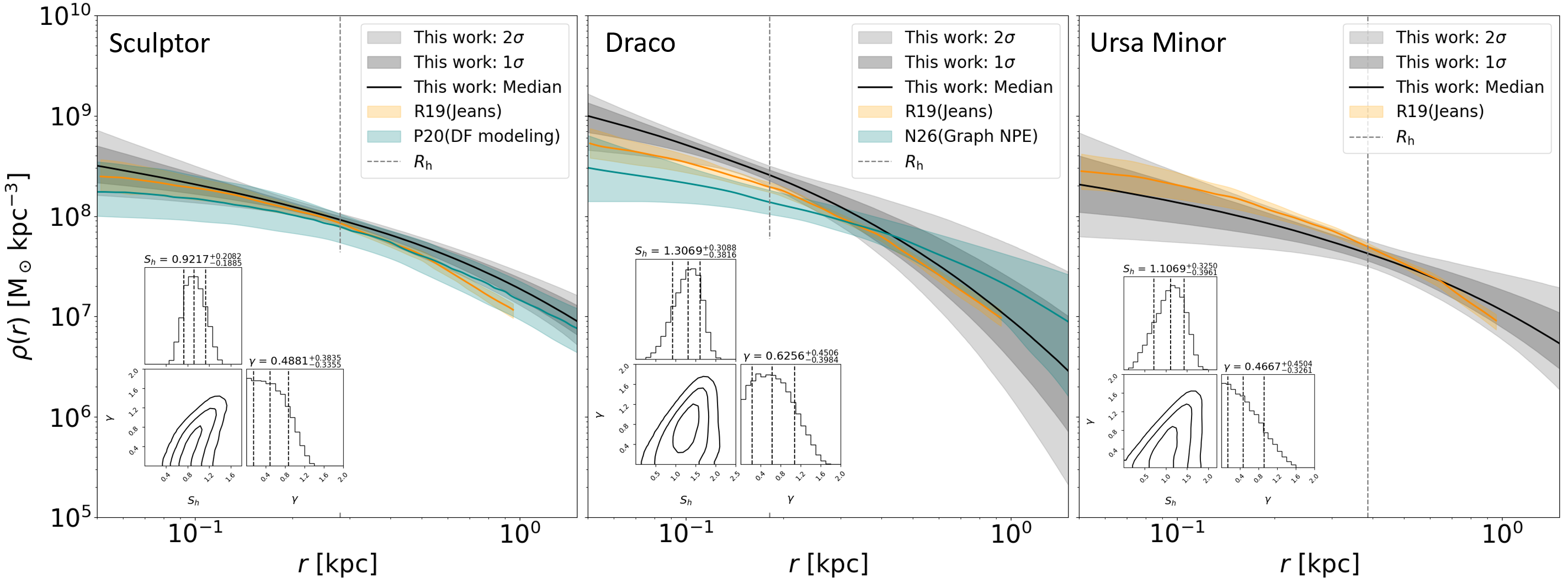}
    \caption{
    Dark matter density profiles inferred for Sculptor, Draco, and Ursa Minor using recent kinematic data sets. The black solid lines show the posterior medians from this work, with the dark and light gray regions indicating the $1\sigma$ and $2\sigma$ credible intervals, respectively. The colored curves show previous results from R19, P20, and N26, with their quoted $1\sigma$ uncertainties. The vertical dashed lines indicate the half-light radii. The inset panels show the posterior distributions of $S_h$ and $\gamma$.
    }
    \label{3dSph_DMprofile}
\end{figure*}

\subsubsection{Dark Matter Density Profile}
Figure~\ref{3dSph_DMprofile} shows the inferred dark matter density profiles for the three galaxies. The black solid lines and gray shaded regions show the posterior medians and the $1\sigma$ and $2\sigma$ credible intervals obtained in this work, respectively. The colored curves show the results from previous studies, for which we display the $1\sigma$ intervals where available. The vertical dashed lines indicate the half-light radii.

For Sculptor, our posterior favors a relatively shallow central density profile, with a median inner slope of $\gamma=0.49^{+0.38}_{-0.34}$ and $S_h=0.92^{+0.21}_{-0.19}$. However, our mock tests show that the recovered \(\gamma\) can be biased when the underlying halo has a large scale radius because such solutions are suppressed by the adopted prior (as discussed in Section \ref{first_discussion}). Therefore, the apparent preference for a shallow \(\gamma\) does not by itself provide a robust determination of the asymptotic inner slope. The inferred density profile is consistent with those obtained by R19 and P20 over the radial range where the results overlap, although our model yields a higher density in the central region. The agreement around the half-light radius is particularly notable, where the different analyses give similar density slopes despite their different modeling approaches.

For Draco, we obtain a relatively steep density slope at the half-light radius, $S_h = 1.31^{+0.31}_{-0.38}$, while the posterior for the inner slope peaks around $\gamma = 0.63^{+0.46}_{-0.40}$. The inferred profile is also denser than the results of R19 and N26 over the range $r \lesssim R_{\rm h}$. In particular, N26 exhibits a markedly different behavior in the inner region. One of the possible reasons is that N26 extends its prior on the inner slope to negative values ($\gamma \in [-1, 2]$), allowing for density profiles that decrease toward the center. This prior specification avoids the boundary truncation at $\gamma = 0$ and naturally shifts their posterior toward lower central densities compared to our framework, which enforces non-negative density gradients.

For Ursa Minor, we compare our result only with R19. We obtain $S_h = 1.11^{+0.33}_{-0.40}$, while the posterior median of the inner slope is $\gamma = 0.47^{+0.45}_{-0.33}$. The density profile is overall consistent with R19, although our credible interval is considerably wider. Indeed, R19 consistently exhibits narrower credible intervals than the other method in all cases. This tight constraint in R19 likely stems from modeling assumptions inherent to their Jeans-based framework (GravSphere). R19 enforces a smooth, four-parameter functional form for the velocity anisotropy and connects the dark matter mass profile using power-law segments across five discrete radial nodes. This parametric simplification restricts the flexibility of the orbital configuration compared to our non-parametric orbit-superposition approach within the framework of Schwarzschild modeling.

The precision of $S_h$ varies between the three galaxies. Sculptor, with approximately 1600 member stars in the adopted sample, gives an uncertainty of about $0.2$ in $S_h$. Draco and Ursa Minor have smaller kinematic samples, of roughly 700 and 500 member stars, respectively, and the uncertainty in $S_h$ increases to about $0.35$. This trend suggests that the number of member stars is an important factor in determining the precision of the inferred density slope. The difference in precision, however, should not be attributed to sample size alone, since the spatial coverage, velocity uncertainties, and intrinsic kinematic structure also differ between the galaxies.

\begin{figure*}
    \centering
    \includegraphics[width=\textwidth]{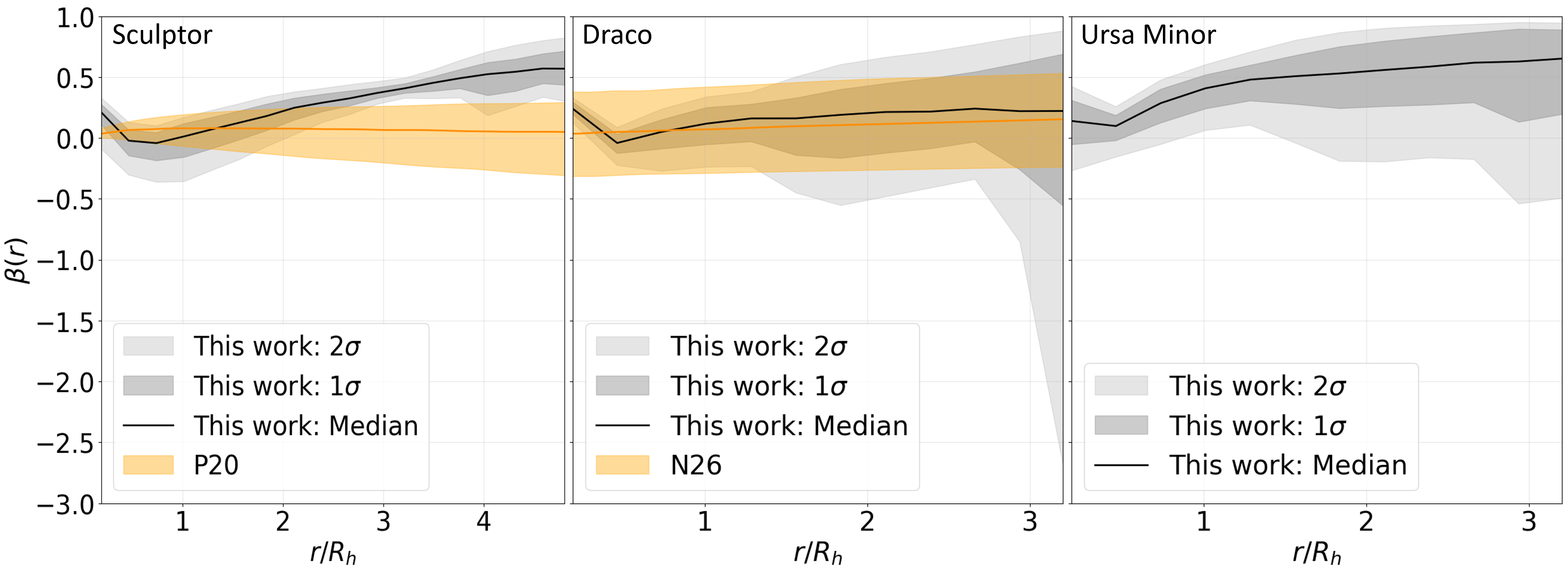}
    \caption{
    Posterior velocity anisotropy profiles of Sculptor, Draco, and Ursa Minor. The black solid lines show the posterior medians from this work, with the dark and light gray regions indicating the $1\sigma$ and $2\sigma$ credible intervals, respectively. The orange curves show the most recently available results from previous studies. The radius is normalized by the projected half-light radius.
    }
    \label{3dSph_beta}
\end{figure*}

\subsubsection{Velocity Anisotropy Profile}
Figure~\ref{3dSph_beta} shows the inferred velocity anisotropy profiles for Sculptor, Draco, and Ursa Minor. The horizontal axis is normalized by the projected half-light radius, as in Figure~\ref{3dSph_DMprofile}. The black solid lines and gray shaded regions show the posterior medians and the $1\sigma$ and $2\sigma$ credible intervals from this work, respectively. For comparison, we show the most recently available result for each galaxy using the orange curves. We do not show a comparison result for Ursa Minor, as we do not find a directly comparable recent constraint on its velocity anisotropy profile.

For Sculptor, our inferred anisotropy profile is better constrained at large radii than the result obtained from the W09 data set in Figure~\ref{beta_W2009}. This improvement is mainly due to the spatial coverage of the adopted kinematic sample. The \citet{2023yCat..36750049T} catalog adopted here includes member stars at larger projected radii, providing additional kinematic information in the outer parts of the galaxy. The posterior median is approximately isotropic near the center and becomes increasingly radially biased toward larger radii. This behavior resembles the general form of an Osipkov--Merritt anisotropy profile (\citealt{1979SvAL....5...42O}; \citealt{1985AJ.....90.1027M}), although no specific functional form for $\beta(r)$ is imposed in our modeling. In contrast, P20 finds a profile that remains close to isotropic over the radial range shown here. This is likely because P20 adopts a parametric action-based distribution function $f(\mathbf{J})$ in which the velocity anisotropy is structurally governed by a single parameter. As discussed in P20, such a simplified analytical form restricts the functional degrees of freedom of the anisotropy profile, making it difficult to capture steep radial transitions in $\beta(r)$ and inherently favoring a nearly constant anisotropy profile.

For Draco, the median anisotropy profile from our model is also close to isotropic over most of the radial range. The result is broadly consistent with N26 at the level of the median profile, although the uncertainty in our inference increases substantially toward the outer region, where the kinematic sample becomes sparse. Interestingly, the agreement in the anisotropy profiles does not correspond to an equally close agreement in the inferred dark matter density profiles shown in Figure~\ref{3dSph_DMprofile}. In particular, the density profiles differ noticeably in the central region, despite the broadly similar anisotropy profiles. This suggests that the difference between the two analyses cannot be attributed solely to the inferred velocity anisotropy.

For Ursa Minor, the inferred anisotropy profile is approximately isotropic near the center and becomes increasingly radially biased at larger radii, again resembling an Osipkov--Merritt-like profile. The outer profile is reasonably well constrained within the radial range covered by the data, although the uncertainty increases toward the largest radii.

\section{Discussion}
\subsection{Prior Sensitivity of the Halo Parameters}
\label{first_discussion}
Earlier Schwarzschild modeling by \citet{2013A&A...558A..35B} showed that the dark matter mass and logarithmic density slope can be constrained nearly independently of the adopted halo density model around $r_{-3}$, where the logarithmic slope of the stellar density profile is $-3$. However, finite-radius quantities are not necessarily insensitive to modeling assumptions in general. Using GravSphere mock tests, \citet{2018MNRAS.481..860R} found that the logarithmic density slope at $150\,\mathrm{pc}$ remained sensitive to the adopted prior, whereas the density amplitude at the same radius was considerably more robust. Our analysis addresses a complementary question by examining the prior sensitivity of the mass and density slope specifically around the stellar half-light radius within a Schwarzschild framework.

As shown in Figure~\ref{pdf_prior_difference}, the enclosed mass $M_h$ and the logarithmic density slope at the projected half-light radius $S_h$ are insensitive to the choice of prior considered here, whereas the posterior distributions of the density-profile parameters $(\rho_0,r_0,\gamma,\kappa)$ differ substantially. In particular, although the posterior medians of $\gamma$ under the two priors are not dramatically different, their posterior distributions have markedly different shapes. This makes $\gamma$ alone an unreliable indicator of whether a dark halo has a core or a cusp.

The different levels of sensitivity to the prior can be understood from the physical meaning of these parameters. The inner and outer slopes, $\gamma$ and $\kappa$, describe the asymptotic behavior of the density profile as $r\rightarrow0$ and $r\rightarrow\infty$, respectively. Neither limit is directly probed by the observed stellar kinematics, which cover only a finite radial range. The scale radius $r_0$ is also poorly constrained when it lies well inside or outside the region containing most of the observed stars. In these regimes, changing $r_0$ can be compensated for by changes in the other density-profile parameters, producing similar kinematic predictions and little difference in likelihood. The posterior distributions of these parameters consequently retain a strong dependence on their priors. The characteristic density $\rho_0$ is strongly degenerate with the other halo parameters, so its posterior distribution changes along with theirs.

By contrast, $M_h$ and $S_h$ characterize the mass distribution and local density slope around the half-light radius, where the stellar kinematic data provide the most direct constraints. Their posterior distributions therefore remain relatively stable under changes in the prior. This distinction between finite-radius quantities and parameters describing the asymptotic density profile is central to understanding what can be inferred from the available data.

The difference between the two priors for constraining $r_0$ can be understood more quantitatively by examining their induced probability densities for $r_0$. If a parameter $X$ has prior density $p(X)$, conservation of probability under a change of variables gives
\begin{equation}
    p(r_0) = p(X)
    \left|\frac{dX}{dr_0}\right|.
\end{equation}
For Prior~1, which is uniform in $S_h$, and Prior~2, which is uniform in $\log_{10}r_0$, the induced prior densities are, respectively,
\begin{equation}
    \label{jacobian_1}
    p(r_0) \propto
    \begin{cases}
        \dfrac{(\kappa-\gamma)R_h}{(R_h+r_0)^2}, & \text{for Prior 1,} \\[10pt]
        \dfrac{1}{r_0}, & \text{for Prior 2,}
    \end{cases}
\end{equation}
where factors independent of $r_0$ have been omitted. For $r_0\gg R_h$, Prior~1 therefore gives $p(r_0)\propto r_0^{-2}$, whereas Prior~2 gives $p(r_0)\propto r_0^{-1}$. Prior~1 thus assigns progressively less probability to large scale radii than Prior~2 does.

This difference is reflected in the posterior distributions shown in Figure~\ref{pdf_prior_difference}: Prior~2 allows larger values of $r_0$ than Prior~1. The associated changes in $r_0$ are accompanied by changes in $\rho_0$, $\gamma$, and $\kappa$, owing to the degeneracies among the density-profile parameters. The resulting differences in their posterior distributions are therefore not simply a matter of statistical uncertainty; they also reflect how the prior weights different regions of the degenerate parameter space.

Taken together, our results are consistent with the earlier conclusion that dark matter properties defined at finite radii can be more robustly constrained than asymptotic halo parameters, and further show that the robustness of $M_h$ and $S_h$ persists under the two different prior weightings considered here.

\subsection{The Role of Higher-order Velocity Moments and Velocity Anisotropy}
\label{second_discussion}

\begin{figure*}
    \centering
    \includegraphics[width=\textwidth]{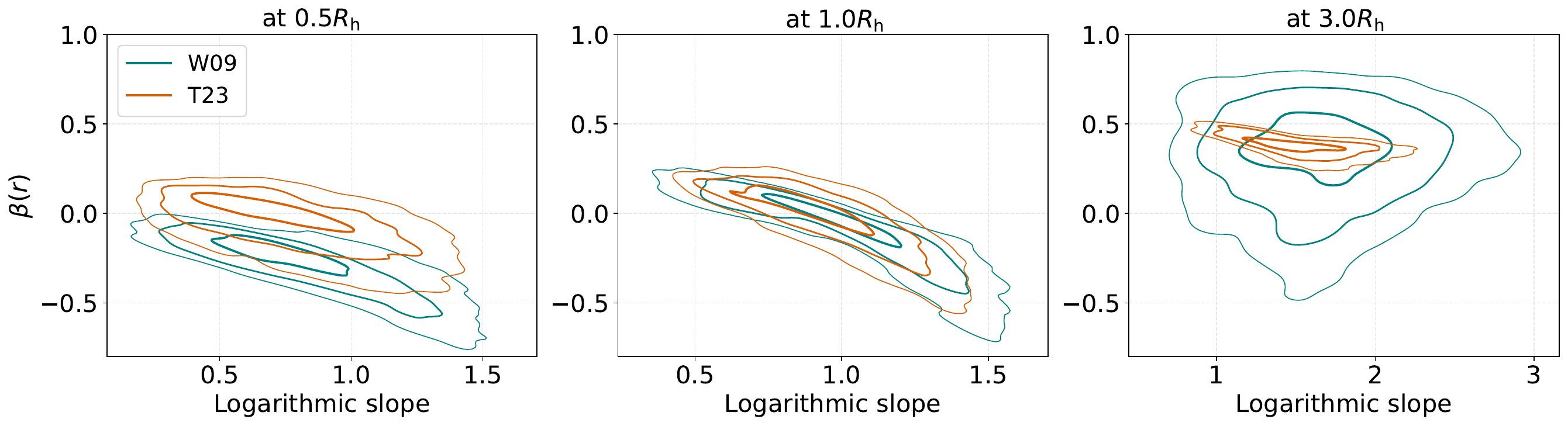}
    \caption{Joint posterior distributions of the logarithmic dark matter density slope and the velocity anisotropy $\beta$ for Sculptor, based on the W09 (green) and T23 (orange) data sets. From left to right, the panels show the slope and anisotropy evaluated at $R=0.5R_h$, $R_h$, and $3R_h$, respectively, where $R_h$ is the projected stellar half-light radius. The contours enclose the $1\sigma$, $2\sigma$, and $3\sigma$ credible regions.}
    \label{slope_anisotropy}
\end{figure*}

As shown in Figure~\ref{beta_W2009}, for Sculptor, our analysis yields a more radially biased velocity anisotropy than that obtained by B13. As discussed in Section~\ref{result1}, this difference is likely related to the different kurtosis profiles inferred by the two analyses. For a velocity distribution with kurtosis greater than three, the high-velocity tails are more prominent than in a Gaussian distribution. Conversely, a kurtosis below three indicates less weight in the tails and a flatter distribution near its peak. Although the relation between kurtosis and velocity anisotropy is not strictly one-to-one due to its dependence on the gravitational potential and distribution function, radial anisotropy ($\beta>0$) generally tends to produce a LOSVD with higher kurtosis for a fixed density profile, whereas tangential anisotropy ($\beta<0$) lowers it, except in the central region (\citealt{2025ApJ...982..167W}; see their Figure~11).

This dependence of the inferred $\beta$ on the kurtosis of the LOSVD suggests that the different result from B13 for $\beta$ stems from the different analysis for the derivation of the LOSVD from the observational data. B13 fits the binned second- and fourth-order velocity moments directly. As shown in Figure~\ref{sigmalos_and_kurtosis}, many of the kurtosis measurements for the W09 data lie below three. Directly fitting these binned measurements may therefore lead to a lower inferred kurtosis. In addition, B13 adds the contributions from the residuals of the two moments directly to the $\chi^2$. Since the second- and fourth-order moments are correlated, treating their residuals as independent can effectively give more weight to the fourth-order information than is justified by the amount of independent information in the data. These factors may explain why B13 inferred a strongly tangential anisotropy profile, with $\beta\sim-0.7$. However, estimates of fourth-order moments from individual radial bins are particularly sensitive to the number of stars and to stars in the tails of the velocity distribution. The observed binned moments may therefore be affected by statistical fluctuations. 

To reduce this sensitivity, our MDN-based approach estimates the velocity distribution from the unbinned data before calculating its moments, yielding smoother velocity moment profiles. Consequently, our inferred kurtosis profile is higher than that obtained by B13, leading to a more radially biased velocity anisotropy profile.

\subsection{Degeneracy between the Density Slope and Velocity Anisotropy}
The impact of velocity anisotropy on the inferred dark matter density profile is illustrated in Figure~\ref{slope_anisotropy}. The figure shows the posterior distributions of the logarithmic density slope and velocity anisotropy at $R=0.5R_h$, $R_h$, and $3R_h$ for Sculptor, based on the W09 and T23 data sets. The green and orange contours show the results for W09 and T23, respectively, with the contours enclosing the $1\sigma$, $2\sigma$, and $3\sigma$ credible regions. As shown in the left and middle panels, the slope and anisotropy exhibit a clear negative correlation around the half-light radius: models with more tangentially biased orbits tend to favor steeper density profiles, whereas models with more radially biased orbits tend to favor shallower profiles. A similar correlation is evident at $0.5R_h$, although the allowed parameter ranges differ between the two data sets. As shown in Figure~\ref{logslope_W2009}, B13 obtains a steeper density slope around the half-light radius. This may be related to its strongly tangential orbital structure, which shifts the inferred slope along the degeneracy. At $3R_h$, the correlation between slope and anisotropy is less pronounced for both data sets. Moreover, the T23 contours are substantially narrower than those for W09, indicating that the additional outer stellar members improve the constraints on the outer anisotropy. This suggests that larger samples of outer stars may allow the velocity anisotropy to be constrained more precisely, with less sensitivity to the slope--anisotropy degeneracy. Better constraints on the outer velocity anisotropy may help characterize the orbital structure of dwarf galaxies and, when combined with other observational constraints and simulations, provide insight into their formation and tidal evolution.

These results connect the treatment of higher-order velocity moments directly to the inference of dark matter density profiles (e.g., \citealt{2025ApJ...982..167W}). When only line-of-sight velocities are available, the kurtosis provides information that helps distinguish between different orbital structures. If this information is estimated unreliably or its statistical dependence on the second moment is not properly accounted for, the inferred anisotropy can shift, carrying the density slope with it through the slope--anisotropy degeneracy. A stable estimate of the kurtosis and a statistically consistent treatment of its covariance with the second moment are therefore important for determining both the orbital structure and the inner density profile of dwarf spheroidal galaxies.

\begin{figure*}
    \centering
    \includegraphics[width=\textwidth]{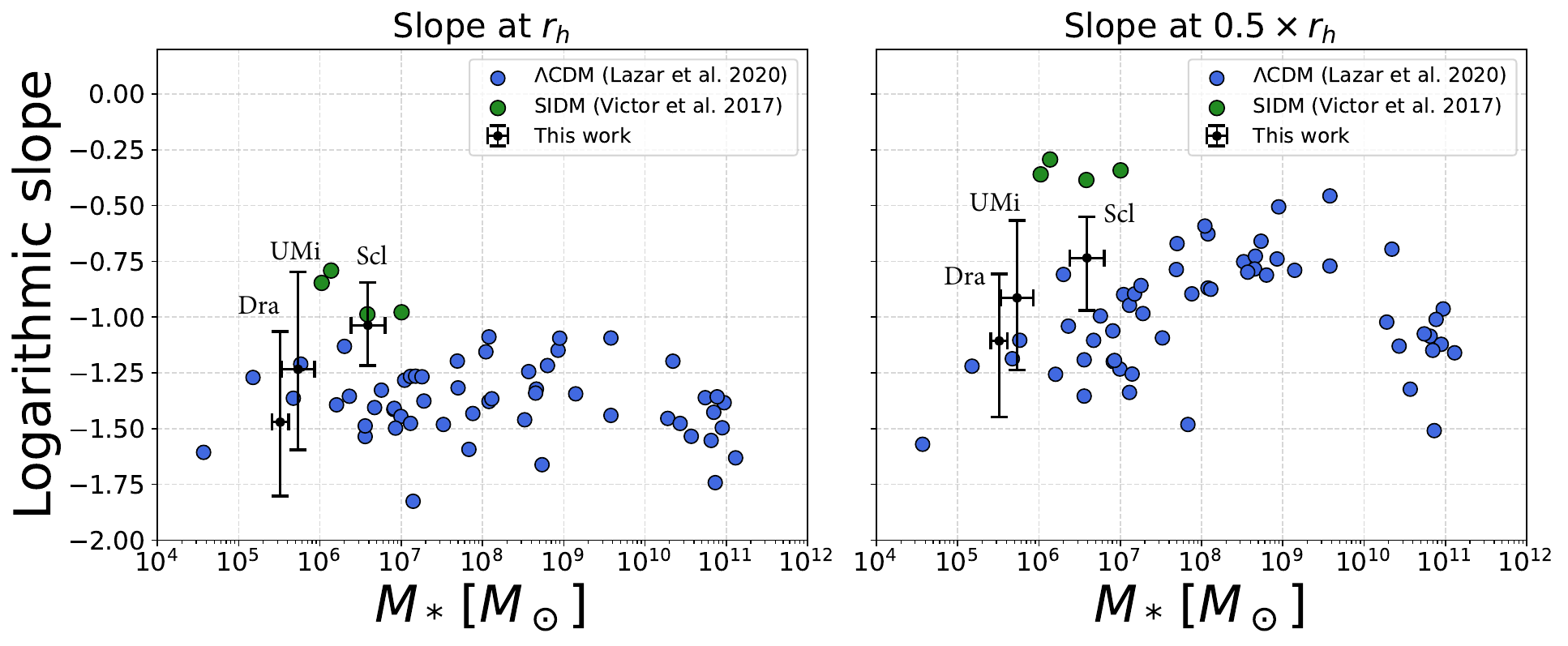}
    \caption{
    Logarithmic dark matter density slope as a function of stellar mass $M_*$. The left and right panels show the slope evaluated at the three-dimensional half-light radius $r_h$ and at $0.5r_h$, respectively. Blue and green points show the FIRE-2 $\Lambda$CDM simulations of \citet{2020MNRAS.497.2393L} and the SIDM simulations with $\sigma/m=1\,{\rm cm^2\,g^{-1}}$ from \citet{2017MNRAS.472.2945R}, respectively. Black points with error bars show the results of this work for Sculptor, Draco, and Ursa Minor, where the error bars indicate the $1\sigma$ credible intervals. Stellar masses are taken from \citet{2013ApJ...779..102K}. These two panels suggest that the baryonic feedback effect on the $\Lambda$CDM density profile is more prominent at $0.5r_h$, at which we can compare with the dark-matter density slopes derived from the dSphs.
    }
    \label{stellarmass_logslope}
\end{figure*}

\subsection{Comparing the Inferred Density Slopes with Cosmological Simulations}
\label{discussion_simulation}
A central question in studies of the core--cusp problem is whether the inner dark matter density profiles inferred from observations are consistent with the predictions of cosmological simulations that include baryonic physics. Previous studies have already demonstrated the utility of finite-radius dark matter constraints for comparisons with cosmological simulations. In particular, \citet{2015ApJ...814L..23B} compared the enclosed mass and logarithmic density slope near $r_{-3}$ inferred for Milky Way dSphs with subhalos from dark-matter-only simulations, showing that the density slope provides information beyond the enclosed mass alone. Here, we extend this line of work using more recent hydrodynamical cosmological simulations (FIRE-2 for $\Lambda$CDM: \citealt{2020MNRAS.497.2393L}) that include baryonic physics, as well as simulations of alternative dark matter models (\citealt{2017MNRAS.472.2945R} for Self-Interacting Dark Matter models: SIDM).

In our analysis, the asymptotic inner slope $\gamma$ is not necessarily the most useful quantity for such a comparison. The limit $r\rightarrow0$ is not directly probed by the stellar kinematics and cannot be resolved directly in cosmological simulations with finite spatial resolution. Moreover, as shown in Figure~\ref{pdf_prior_difference}, $\gamma$ is sensitive to the choice of prior in our dynamical modeling. We therefore compare the logarithmic density slope at finite radii. 

For comparison with the three-dimensional density profiles of the simulations, we evaluate the slopes at the three-dimensional stellar half-light radius, $r_h$, and at $0.5r_h$. Previous comparisons between observed dwarf galaxies and hydrodynamical simulations have often characterized the inner density profile at a fixed fraction of the virial radius and examined its dependence on the stellar-to-halo mass ratio (e.g., \citealt{2017ARA&A..55..343B}; \citealt{2020ApJ...904...45H}). For observed dSphs, however, neither the virial radius nor the total halo mass is directly constrained by the stellar kinematic data, and their estimates may rely on empirical stellar-to-halo mass relations such as those inferred from abundance matching, which become increasingly uncertain in the dwarf-galaxy regime (e.g., \citealt{2014ApJ...784L..14B}; \citealt{2015MNRAS.448.2941S}). Our comparison instead uses the stellar mass and the stellar half-light radius, avoiding the need to infer $M_{\rm halo}$ or $R_{\rm vir}$ and thereby reducing the dependence on an additional galaxy--halo connection model.

Figure~\ref{stellarmass_logslope} compares our measurements of the dark matter density slopes with predictions from cosmological simulations as a function of stellar mass. The blue points show the FIRE-2 $\Lambda$CDM simulations of \citet{2020MNRAS.497.2393L}, while the green points show the SIDM simulations with $\sigma/m=1\,{\rm cm^2\,g^{-1}}$ presented by \citet{2017MNRAS.472.2945R}. The black points with error bars show our results for Sculptor, Draco, and Ursa Minor. The stellar masses are taken from \citet{2013ApJ...779..102K}. The left and right panels show the density slopes at $r_h$ and $0.5r_h$, respectively.

At $r=0.5r_h$ (the right panel of Figure~\ref{stellarmass_logslope}), the CDM simulations show a strong dependence of the density slope on stellar mass. In the FIRE-2 simulations, feedback-driven modifications to the dark matter density profile are most pronounced at $M_*\sim10^9$--$10^{10}\,M_\odot$, where substantial cores can form. At lower stellar masses, the effects of baryonic feedback become less pronounced, and the simulated density profiles tend to become steeper. The SIDM simulations shown here tend to exhibit shallower density profiles than the $\Lambda$CDM simulations over the stellar mass range covered by the available SIDM models. In the SIDM model of \citet{2017MNRAS.472.2945R}, self-interactions with $\sigma/m=1\,{\rm cm^2\,g^{-1}}$ redistribute energy among dark matter particles, leading to a shallower central density profile. Our measurements lie broadly within the range spanned by the $\Lambda$CDM simulations. Draco, in particular, lies relatively close to the $\Lambda$CDM predictions in this comparison. These results should not, however, be interpreted as evidence that $\Lambda$CDM is preferred over SIDM for several reasons. First, the SIDM comparison is limited to a single cross section and a small number of simulated systems, and therefore does not represent the full range of possible SIDM predictions. Second, our measurements have substantial uncertainties and cover only three galaxies, while the simulations sample a limited range of galaxy properties and dark matter models. In addition, our dynamical model assumes spherical symmetry, whereas real dwarf galaxies need not be spherical (see Section~\ref{limitations}).

At $r=r_h$ (the left panel), the density slopes in the $\Lambda$CDM simulations show a weaker dependence on stellar mass than at $0.5r_h$. One possible explanation is that, in systems where feedback produces a core, $r_h$ lies outside the region where the density profile is substantially flattened (\citealt{2020MNRAS.497.2393L}, see their Figure~8), resulting in a steeper slope at this radius. The SIDM simulations also exhibit relatively steep slopes around $r_h$, although their slopes appear to remain slightly shallower than those of the $\Lambda$CDM simulations in this small sample. Nevertheless, more precise measurements, enabled by larger spectroscopic samples and also the addition of proper-motion data, may help establish whether finite-radius density slopes, and at which radii, can distinguish between different dark matter models.

\subsection{Limitations and Future Prospects}
\label{limitations}
Our modeling framework has several limitations that should be kept in mind when interpreting the results. First, as in B13, we use the solution of the quadratic programming (QP) problem as a representative set of orbital weights for each set of halo parameters, rather than marginalizing over the orbital weights. Consequently, the uncertainty in the orbital weights is not fully accounted for, and the resulting uncertainties in the velocity anisotropy and dark matter parameters may be underestimated. \citet{2026arXiv260700117Z} reduced this limitation by solving the QP problem 100 times to account for variations in the orbital weights. Nevertheless, rigorous marginalization over the orbital weights remains an important challenge in Schwarzschild modeling.

Second, we assume spherical symmetry for both the stellar distribution and the dark matter halo, which may introduce finite systematic biases. Observations show that the stellar distributions of dwarf spheroidal galaxies can be flattened (e.g., \citealt{2018ApJ...860...66M}), while simulations indicate that dark matter halos can be triaxial rather than spherical in both $\Lambda$CDM and SIDM models (\citealt{2002ApJ...574..538J}; \citealt{2006MNRAS.367.1781A}; \citealt{2017MNRAS.472.2945R}). \citet{2018MNRAS.476.2918K} investigated spherical Schwarzschild modeling of a simulated axisymmetric dwarf galaxy and found that the inferred velocity anisotropy can depend strongly on the viewing direction. In their tests, when the galaxy was viewed along its minor axis, the anisotropy was generally underestimated, even when the overall density profile was known. Such systematic effects may also propagate into the inferred dark matter density slopes. Extending our framework to allow for axisymmetry is therefore an important next step.

Third, the available stellar sample size remains small for a Schwarzschild model with many degrees of freedom, limiting the constraints that can be placed on the halo properties and orbital structure. Increasing the number of member stars is therefore an important priority, and the ongoing Subaru Prime Focus Spectrograph (PFS) project is expected to make substantial progress in this regard (see \citealt{2026arXiv260409875C}). In particular, samples of more than $\sim1000$ member stars are expected for Draco and Ursa Minor. These larger samples should reduce the statistical uncertainties on the density slopes and may improve our ability to distinguish between dark matter models through comparisons with cosmological simulations, as in Figure~\ref{stellarmass_logslope}. PFS will also improve the sampling of the outer regions, where the velocity anisotropy is currently less well constrained. Better constraints on the outer anisotropy may provide additional insight into the orbital structure and formation histories of dwarf galaxies, particularly when combined with other observational constraints and simulations.

\section{CONCLUSIONS}
\label{sec:conclusions}

In this work, we have developed a spherical Schwarzschild model with an MDN-based treatment of the LOSVD to constrain the dark matter density profiles and stellar orbital structures of Milky Way dwarf spheroidal galaxies. The LOSVD is represented by a Gaussian mixture whose parameters are predicted as smooth functions of projected radius by a mixture density network trained on the unbinned stellar velocities. From the resulting LOSVD, we have derived smooth second- and fourth-order velocity-moment profiles and use them as kinematic constraints in the Schwarzschild modeling, accounting for both their covariance and the correlations introduced by the smoothing procedure. We have further combined the Schwarzschild calculation with MCMC sampling of the dark matter halo parameters. Tests with four mock data sets show that the dark matter density profiles and their slopes around the half-light radius are generally recovered well, although the inferred velocity anisotropy can show systematic deviations in some radial ranges.

Applying our method to Sculptor, we have found that the enclosed mass and logarithmic density slope at the projected half-light radius, $M_h$ and $S_h$, remain nearly unchanged between the two halo priors considered here, whereas the asymptotic inner slope $\gamma$ and scale radius $r_0$ show substantially stronger prior dependence. This behavior is consistent with previous studies showing that dark matter properties can be constrained more robustly at finite radii than in the asymptotic inner limit. Our analysis further demonstrates that, within the present Schwarzschild framework, the slope evaluated around the half-light radius is considerably less sensitive to the adopted prior weighting than the asymptotic slope $\gamma$.

Using recent spectroscopic data sets, we have obtained $S_h=0.92^{+0.21}_{-0.19}$, $1.31^{+0.31}_{-0.38}$, and $1.11^{+0.33}_{-0.40}$ for Sculptor, Draco, and Ursa Minor, respectively. The posterior median anisotropy profiles show different radial behavior among the three systems: Sculptor and Ursa Minor are approximately isotropic near the center and become increasingly radially biased toward larger radii, whereas Draco remains close to isotropic over most of the radial range probed by the data. The larger radial coverage of the recent Sculptor sample provides narrower posterior for its outer anisotropy than the W09 data set. In view of the systematic deviations seen in the mock tests, however, detailed features of the recovered anisotropy profiles should be interpreted with caution.

Our comparison with the previous Schwarzschild analysis of Sculptor also highlights the importance of higher-order velocity information. Differences in the inferred fourth-order moment are associated with different recovered anisotropy profiles, which in turn correlate with the inferred dark matter density slope through the density-slope--anisotropy degeneracy. Stable estimation of the LOSVD shape and a statistically consistent treatment of the covariance between the second- and fourth-order moments are therefore important for constraining both the stellar orbital structure and the dark matter density profile.

We have further compared the inferred dark matter density slopes at the three-dimensional stellar half-light radius, $r_h$, and at $0.5r_h$ with predictions from $\Lambda$CDM and SIDM cosmological simulations including baryonic feedback as a function of stellar mass. This comparison is expressed in terms of $M_*$ and $r_h$ tied to the observed stellar distribution, avoiding the need to introduce halo-scale quantities such as the total halo mass $M_{\rm halo}$ or the virial radius $R_{\rm vir}$ that are not directly constrained by the stellar data. At $0.5r_h$, the inferred slopes of Sculptor, Draco, and Ursa Minor lie broadly within the range spanned by the FIRE-2 $\Lambda$CDM simulations. The present uncertainties, the small number of observed systems, and the limited set of available SIDM simulations, however, do not yet allow a robust distinction between dark matter models. Larger spectroscopic samples will be essential for determining whether the dependence of the finite-radius density slope on stellar mass can provide a useful discriminator between competing dark matter and galaxy-formation scenarios.

Several limitations remain. Our current model assumes spherical symmetry and uses a single optimized set of orbital weights for each gravitational potential rather than fully marginalizing over the orbital-weight distribution. Extending the framework to axisymmetric stellar systems, improving the statistical treatment of the orbital weights, and incorporating larger kinematic samples and proper-motion information are therefore important next steps. Upcoming spectroscopic surveys such as Subaru/PFS will be particularly valuable for increasing the number of member stars and improving the sampling of the outer regions, where the orbital structure is currently less well constrained.

Overall, our results support characterizing the inner dark matter structure of dwarf galaxies using observable quantities evaluated over the radial range directly probed by the stellar kinematics, rather than relying on the asymptotic inner slope alone. Combined with comparisons to cosmological simulations based on stellar mass and stellar size scales, flexible orbit-based modeling with robust treatment of higher-order kinematic information provides a promising route toward testing the origin of the inner dark matter structure of dwarf galaxies.

\begin{acknowledgments}
We are grateful to Dafa Wardana and Kohei Hayashi for helpful discussions on the dynamical analysis of Milky Way dwarf spheroidal galaxies. ChatGPT (OpenAI) and Gemini (Google) were used to assist with coding, debugging, proofreading, and language refinement during the preparation of this work. The authors reviewed and verified all resulting content and take full responsibility for the final manuscript. This work was supported in part by the JSPS KAKENHI Grant (No. JP24K00669, JP25K24687 for M.C.; No. 23K13098, 25KJ0017 for S.H.). 
\end{acknowledgments}


\appendix
\renewcommand{\thefigure}{\thesection\arabic{figure}}
\setcounter{figure}{0}

\section{Stellar Surface Density Fitting}
\label{stellar_fitting}
To determine the target stellar surface density distribution to be reproduced by the quadratic programming (QP) optimization, we fit an analytical profile (equation \eqref{stellar_surface_density}) to the observed two-dimensional projected coordinates of the stars. To construct the likelihood function for this fitting, we directly utilize the individual stellar coordinates $R_i$ without binning the data. The probability density function $P(R)$ of finding a star at a projected radius $R$ is defined by weighting the surface density $\Sigma(R)$ by the polar coordinate area element $2\pi R$ and normalizing it over the entire observational domain:
\begin{equation}
    P(R_i | \boldsymbol{\theta}) = \frac{2\pi R_i \Sigma(R_i; \boldsymbol{\theta})}{\int_0^{R_{\text{max}}} 2\pi R \Sigma(R; \boldsymbol{\theta}) dR}
\end{equation}
where $\boldsymbol{\theta} = \{R_s, \alpha_*, \beta_*, \gamma_*\}$ represents the set of parameters to be estimated. The integral in the denominator, denoted as $N_{\text{norm}}(\boldsymbol{\theta})$, serves as the normalization to ensure that the total probability integrates to unity. The upper limit $R_{\text{max}}$ is set to the maximum projected radius of the stellar sample.

For a sample containing $N$ stars, the unbinned log-likelihood function $\ln \mathcal{L}(\boldsymbol{\theta})$ is formulated as the sum of the individual log-probabilities:
\begin{equation}
    \ln \mathcal{L}(\boldsymbol{\theta}) = \sum_{i=1}^{N} \left[ \ln(2\pi R_i) + \ln \Sigma(R_i; \boldsymbol{\theta}) - \ln N_{\text{norm}}(\boldsymbol{\theta}) \right]
\end{equation}
Using an unbinned likelihood avoids information loss caused by data grouping and eliminates systematic biases—such as artificial shifts in parameter estimates and the distortion of uncertainty bounds—associated with the choice of bin widths and boundary positions, thereby maximizing the statistical information extracted from the data set.

We utilize Bayes' theorem to estimate the joint posterior probability distribution $p(\boldsymbol{\theta} | \{R_i\})$ by combining the likelihood with the prior probability distribution $p(\boldsymbol{\theta})$:
\begin{equation}
    p(\boldsymbol{\theta} | \{R_i\}) \propto \mathcal{L}(\boldsymbol{\theta}) p(\boldsymbol{\theta})
\end{equation}
We assume independent uniform prior distributions for the parameters with the following physically motivated ranges:
\begin{itemize}
    \item $R_s \in [10, 5000] \text{ [pc]}$: A broad interval that comfortably encompasses the typical size of dwarf spheroidal galaxies.
    \item $\alpha_* \in [0.5, 10.0]$: Regulates the sharpness of the transition between the inner and outer regions.
    \item $\beta_* \in [2.1, 10.0]$: The outer density slope. A value of $\beta_* > 2$ is mathematically required to ensure that the integrated total mass in the two-dimensional projection ($\int R \Sigma dR$) converges at infinity.
    \item $\gamma_* \in [0.0, 1.99]$: The inner density slope. To prevent the cumulative mass from diverging near the center ($R \to 0$), the slope is restricted to $\gamma_* < 2$.
\end{itemize}
These bounds ensure physical consistency (i.e., finite stellar mass) while providing sufficient flexibility to cover various stellar configurations.

These parameters are often highly degenerate, resulting in narrow, curving degeneracies within the parameter space. Local optimization algorithms, such as gradient-based methods, easily get trapped in local extrema under such conditions, failing to locate the global maximum. To explore this parameter space efficiently, we employ the No-U-Turn Sampler (\citealt{JMLR:v15:hoffman14a}), an advanced Hamiltonian Monte Carlo (HMC) algorithm, as implemented in NumPyro (\citealt{JMLR:v20:18-403}) with JAX (\citealt{jax2018github}). HMC suppresses diffusive random-walk behavior typical of standard Metropolis--Hastings samplers and efficiently traces complex multidimensional posterior distributions, providing accurate assessments of parameter uncertainties and their correlations.

To identify the surface density profile used for the subsequent Schwarzschild modeling, we adopt a two-step approach to locate the maximum a posteriori (MAP) estimate in the multidimensional parameter space. Simple marginalized estimators, such as the median or marginal mode of each parameter from the MCMC chain, can yield parameter combinations with low joint posterior probability when strong parameter correlations are present.
We first evaluate the log-likelihood for all samples generated by the MCMC search to identify the point with the highest posterior density. This point serves as an excellent initial guess near the global optimum. Finally, we use this configuration as the starting point for a local optimization using the L-BFGS-B algorithm (\citealt{10.1145/279232.279236}). This step eliminates sampling resolution limits imposed by a finite MCMC sample size and yields the precise MAP estimate in the continuous parameter space. The surface density $\Sigma_{\rm fit}$ obtained by substituting the parameters derived from this fitting is used in equations \eqref{mu2}, \eqref{mu4}.

\section{The Detail of the LOSVD Modeling via Mixture Density Network}
\label{MDN_detail}
\subsection{Neural Network Architecture and Orthogonal Parameterization}
\label{subsec:mdn_arch}

To model the Gaussian Mixture Model (GMM) parameters as smooth continuous functions of projected radius $R$, we implement a Mixture Density Network (MDN) based on a multi-layer perceptron (MLP). The network maps the spatial location of individual stars to the parameters governing the local LOSVD.

\subsubsection{Input Feature Standardization}
To effectively capture the large dynamical range of projected radii spanning from the inner region to the outer halo outskirts, we define two complementary spatial features:
\begin{align}
    x_1 &= \log_{10}\left(\frac{R}{\text{pc}}\right), \\
    x_2 &= \frac{1}{R + R_{\mathrm{mild}}},
\end{align}
where $R_{\mathrm{mild}}$ is a soft scaling constant set to $R_{\mathrm{half}}/3$. To prevent feature scale disparities from biasing gradient descent during optimization, both input features are standardized using Z-score normalization evaluated across the entire stellar sample:
\begin{equation}
    x_1' = \frac{x_1 - \mu_{x_1}}{\sigma_{x_1}}, \quad
    x_2' = \frac{x_2 - \mu_{x_2}}{\sigma_{x_2}},
\end{equation}
where $(\mu_{x_1}, \sigma_{x_1})$ and $(\mu_{x_2}, \sigma_{x_2})$ denote the sample means and standard deviations of $x_1$ and $x_2$, respectively. This guarantees that all input signals fed into the network possess zero mean and unit variance.

\subsubsection{Orthogonal Reparameterization Strategy}
Directly predicting the physical dispersion scales and velocity offset $(\sigma_1, \sigma_2, v_0)$ introduces a severe optimization degeneracy: any gradient update attempting to adjust the non-Gaussian shape (kurtosis) unintentionally alters the total second moment $\mu_2$, forcing gradient descent algorithms (e.g., Adam) to bounce around narrow, ill-conditioned valleys in the loss landscape.

To eliminate this scale-shape cross-talk, we decouple the overall velocity dispersion scale from the relative distribution shape. Instead of predicting physical quantities directly, the network is formulated to output a scalar for the total velocity dispersion scale $\sigma_{\mathrm{tot}}$, alongside a set of dimensionless relative shape parameters $(\hat{\sigma}_1, \hat{\sigma}_2, \hat{v}_0)$.

\subsubsection{Network Topology and Output Activation}
The network architecture consists of a multi-layer perceptron with three hidden layers of 64 units each, employing Mish activation functions \citep{DBLP:conf/bmvc/Misra20}. Taking the standardized spatial features $(x_1', x_2')$ as input, the linear output layer produces a 5-dimensional raw vector:
\begin{equation}
    \mathbf{o} = [o_1, o_2, o_3, o_4, o_5]^T.
\end{equation}
These raw values are transformed into the mixing weight $\pi_1$, total scale $\sigma_{\mathrm{tot}}$, and dimensionless shape parameters via dedicated activation functions:
\begin{align}
    \pi_1 &= \text{sigm}(o_1) \in (0, 1), \\
    \sigma_{\mathrm{tot}} &= \exp(o_2) > 0, \\
    \hat{\sigma}_1 &= \text{softplus}(o_3) > 0, \\
    \hat{\sigma}_2 &= \text{softplus}(o_4) > 0, \\
    \hat{v}_0 &= \text{softplus}(o_5) > 0.
\end{align}
Here, the sigmoid function $\text{sigm}(x) = (1 + e^{-x})^{-1}$ constrains $\pi_1$ to $(0, 1)$, and the exponential function guarantees a positive total velocity scale $\sigma_{\mathrm{tot}}$. For the dimensionless shape parameters $(\hat{\sigma}_1, \hat{\sigma}_2, \hat{v}_0)$, we adopt the smooth softplus function, $\text{softplus}(x) = \ln(1 + e^x)$, rather than an exponential function. Because the derivative of $\text{softplus}(x)$ is naturally bounded by the sigmoid function ($\mathrm{d}\,\text{softplus}(x)/\mathrm{d}x = \text{sigm}(x) \le 1$), it provides a smooth, linear-like response for positive values while effectively preventing exploding gradients during backpropagation.

\subsubsection{Physical Parameter Reconstruction and Scale Invariance}
From the predicted dimensionless parameters, the physical quantities appearing in equation \eqref{GMM_moment} are reconstructed while preserving the predicted scale $\sigma_{\mathrm{tot}}$. We first evaluate the dimensionless shape variance $V_{\mathrm{shape}}$:
\begin{equation}
    V_{\mathrm{shape}} = \pi_1 \hat{\sigma}_1^2 + (1 - \pi_1) \left(\hat{\sigma}_2^2 + \hat{v}_0^2\right).
\end{equation}
We then compute a normalization factor $K$ to scale the relative shape to the target dispersion $\sigma_{\mathrm{tot}}$:
\begin{equation}
    K = \frac{\sigma_{\mathrm{tot}}}{\sqrt{V_{\mathrm{shape}}} + \epsilon},
\end{equation}
where $\epsilon = 10^{-8}$ prevents numerical division by zero. The physical parameters $(\sigma_1, \sigma_2, v_0)$ are derived via:
\begin{equation}
    \sigma_1 = K \hat{\sigma}_1, \quad \sigma_2 = K \hat{\sigma}_2, \quad v_0 = K \hat{v}_0.
\end{equation}
Under this orthogonal transformation, the theoretical second-order velocity moment of the GMM identically satisfies:
\begin{align}
    \mu_{2,\mathrm{GMM}} &= \pi_1 \sigma_1^2 + (1 - \pi_1) \left(\sigma_2^2 + v_0^2\right) \nonumber \\
    &= K^2 \left[ \pi_1 \hat{\sigma}_1^2 + (1 - \pi_1) \left(\hat{\sigma}_2^2 + \hat{v}_0^2\right) \right] \nonumber \\
    &= K^2 V_{\mathrm{shape}} = \sigma_{\mathrm{tot}}^2.
\end{align}
This algebraic identity guarantees that any perturbation of the dimensionless shape parameters $(\hat{\sigma}_1, \hat{\sigma}_2, \hat{v}_0)$ leaves the predicted total velocity dispersion strictly invariant at $\sigma_{\mathrm{tot}}^2$. By reducing the coupling between the velocity scale and shape parameters, this reparameterization improves the conditioning of the negative log-likelihood landscape, enabling more stable and efficient convergence during optimization.

\subsection{Loss Function, Optimization, and Ensemble Training}
\label{subsec:mdn_loss_training}

\subsubsection{Loss Function Formulation}
The network is trained by minimizing a multi-component negative log-likelihood (NLL) loss function:
\begin{equation}
    \mathcal{L}_{\mathrm{MDN}} = \mathcal{L}_{\mathrm{base}} + \alpha_{\mathrm{vip}} \mathcal{L}_{\mathrm{tail}},
\end{equation}
where $\mathcal{L}_{\mathrm{base}}$ evaluates the standard likelihood across all stars, $\mathcal{L}_{\mathrm{tail}}$ penalizes misfits in the high-velocity tails, and $\alpha_{\mathrm{vip}} = 0.3$ controls the relative strength of the tail penalty.

To place greater emphasis on the inner galactic region—which provides critical kinematic constraints for breaking the core-cusp degeneracy—we incorporate a radial spatial weighting function $W(R)$:
\begin{equation}
    W(R) = 1 + A \exp\left[-\left(\frac{R}{R_{\mathrm{core}}}\right)^2\right],
\end{equation}
where $A = 3.0$ and $R_{\mathrm{core}} = R_{\mathrm{half}}/6.0$. Individual stellar velocity measurement uncertainties $e_i$ are explicitly convolved into each Gaussian component by replacing $\sigma_k^2$ with $\sigma_{k,\mathrm{obs}}^2 = \sigma_k^2 + e_i^2$. The base NLL is then given by:
\begin{equation}
    \mathcal{L}_{\mathrm{base}} = -\frac{1}{N} \sum_{i=1}^{N} W(R_i) \ln \left[ p(v_i \mid R_i, e_i) \right],
\end{equation}
where $p(v_i \mid R_i, e_i)$ is the LOSVD described by GMM (equation \eqref{GMM}).

To capture non-Gaussian kurtosis without being dominated by the far-more-numerous central velocity stars, we define the tail-weighted loss $\mathcal{L}_{\mathrm{tail}}$:
\begin{equation}
    \mathcal{L}_{\mathrm{tail}} = -\frac{1}{N} \sum_{i=1}^{N} W(R_i) w_{\mathrm{vip}}(v_i) \ln \left[ p(v_i \mid R_i, e_i, \text{sg}) \right],
\end{equation}
where $w_{\mathrm{vip}}(v_i) = \text{sigm}\left((|v_i| - v_{\mathrm{cut}})/\tau\right)$ is a smooth Soft-VIP (Velocity Importance) weighting function with $v_{\mathrm{cut}} = 2.0$ (in units of the velocity dispersion computed from the full stellar sample, $\sigma_{v,\mathrm{global}}$) and a smoothing scale $\tau = 0.1$. 

Crucially, "$\text{sg}$" denotes a stop-gradient operator ($\text{detach}$) applied to the overall scale $\sigma_{\mathrm{tot}}$ within $\mathcal{L}_{\mathrm{tail}}$. By cutting gradient backpropagation to $\sigma_{\mathrm{tot}}$, $\mathcal{L}_{\mathrm{tail}}$ is strictly constrained to update only the dimensionless shape parameters $(\hat{\sigma}_1, \hat{\sigma}_2, \hat{v}_0)$. This isolates the determination of the global second-order moment $\mu_2$ from high-velocity tail fluctuations while enabling precise fitting of the fourth-order moment $\mu_4$.

\subsubsection{Optimization and Hyperparameters}
Before training, all stellar velocities $v_i$ and measurement errors $e_i$ are normalized by the sample standard deviation $v_{\mathrm{scale}} = \mathrm{std}(v)$. Training is performed using the Adam optimizer (\citealt{DBLP:journals/corr/KingmaB14}) with an initial learning rate $\eta = 10^{-3}$, a weight decay of $\mathcal{O}(10^{-3})$ and a mini-batch size of 512. The learning rate is annealed over $T_{\mathrm{max}} = 400$ epochs using a Cosine Annealing schedule (\citealt{DBLP:conf/iclr/LoshchilovH17}):
\begin{equation}
    \eta_t = \frac{\eta}{2} \left[ 1 + \cos\left( \frac{\pi t}{T_{\mathrm{max}}} \right) \right].
\end{equation}

\subsubsection{Ensemble Learning and Symmetric Trimming}
To suppress single-model variance arising from random weight initialization and shot noise in finite stellar samples, we employ an ensemble of $M = 128$ independent MDN models running in parallel.

The first model ($m = 0$) is trained on the unresampled original data set. For all remaining models ($m = 1, \dots, 127$), bootstrap resamples are generated by drawing, with replacement, the same number of stars as in the original data set, with each model also assigned a distinct random initialization seed.

To filter out atypical optimization outcomes (e.g., rare models trapped in local sub-optima or overfitted instances), we evaluate the final loss performance of each ensemble member by averaging its NLL over the last 20 epochs. We discard models falling below the 2nd percentile or exceeding the 98th percentile of the final loss distribution (a 2\% symmetric trim on both ends).

\section{Mock data test}
\label{mock}
\subsection{Mock Data} 
We generated four mock galaxies using AGAMA \citep{2019MNRAS.482.1525V} to test the performance and robustness of our method under a variety of physically motivated conditions. Mock~1 adopts a cuspy Hernquist profile (\citealt{1990ApJ...356..359H}) and serves as a fiducial benchmark model. Mock~2 employs a dark matter halo with a relatively large scale radius ($r_0/R_{\rm half} = 25$) to verify whether our model can properly estimate the density profile of such an extended halo. Mock~3 represents a cored dark matter halo, allowing us to assess the ability of the method to distinguish between cuspy and cored density profiles. Finally, Mock~4 follows an Osipkov--Merritt (OM) anisotropy profile with an anisotropy scale radius of $r_a = 7 R_h$, providing a test in the presence of a radially varying velocity anisotropy. The central dark matter densities $\rho_0$ were scaled to ensure that all systems remain strongly dark-matter dominated, maintaining a stellar-to-halo mass ratio of $M_* / M_{\rm halo} \sim 10^{-4}$, typical of dwarf spheroidal galaxies. All properties of the mock galaxies are summarized in Table~\ref{tab:mocks}.

For each mock galaxy, we generated $10^5$ stellar tracers. Stars located within a projected radius of $5R_{\rm half}$ were randomly sampled to construct mock catalogs containing $1500$ stars. This sample size reflects the number of member stars expected for nearby dwarf spheroidal galaxies observed with the Prime Focus Spectrograph (PFS). Gaussian line-of-sight velocity uncertainties with a standard deviation of $2~{\rm km\,s^{-1}}$ were added to each star. Note that the MCMC was run with the Prior 1 setting.

\begin{table}
\centering
\caption{Summary of the structural and dynamical parameters for the mock galaxies. The parameters $\rho_0$, $r_0$, $\gamma$, and $\kappa$ define the dark matter halo profile (equation \eqref{DM density}). The central dark matter density $\rho_0$ is set such that $M_* / M_{\rm halo} \sim 10^{-4}$. $\beta(r)$ denotes the velocity anisotropy profile, where OM indicates the Osipkov--Merritt profile with $r_a = 7 R_h$.}
\label{tab:mocks}
\begin{tabular}{lcccc}
\toprule
Parameter & Mock 1 & Mock 2 & Mock 3 & Mock 4 \\
\midrule
$\rho_0\ [M_\odot / \text{pc}^3]$ & $2.4 \times 10^{-2}$ & $6.3 \times 10^{-3}$ & $6.9 \times 10^{-2}$ & $4.6 \times 10^{-2}$ \\
$r_0 / R_{\text{h}}$ & 4 & 25 & 4 & 8 \\
$\gamma$ & 1 & 1 & 0 & 0 \\
$\kappa$ & 4 & 3 & 3 & 3 \\
$\beta(r)$ & $-0.5$ & 0 & 0 & OM \\
$\rho_*(r)$ & Plummer & Plummer & Plummer & Plummer \\
\bottomrule
\end{tabular}
\end{table}

\subsection{The Recovery of Velocity Moments Using the MDN}
\begin{figure*}
    \centering
    \includegraphics[width=\textwidth]{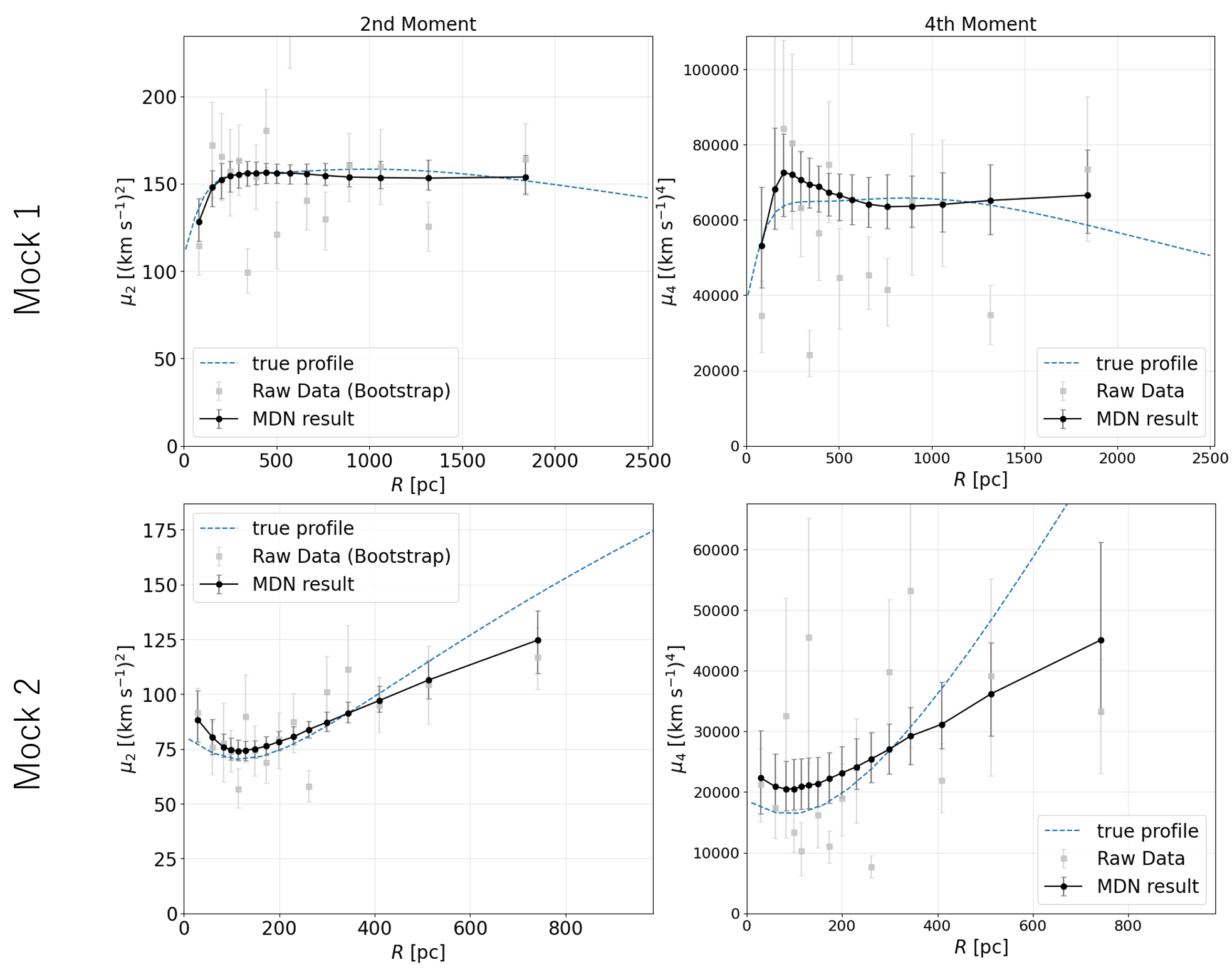}
    \caption{Comparison of the velocity moments estimated by the MDN with the true profiles in mock data sets. The left and right columns show the second- and fourth-order velocity moments, respectively. The gray points show the moments measured directly from the mock data, with uncertainties estimated by bootstrap resampling. The black points show the results obtained from the MDN, and the blue dashed curves indicate the true profiles used to generate the mock data. The upper and lower rows show Mock 1 and Mock 2, respectively. Each mock data set contains $1500$ member stars. The two mocks represent different radial behaviors of the true second moment, with Mock 1 decreasing and Mock 2 increasing toward the center.
    }
    \label{MDN}
\end{figure*}

In this section, we test whether the MDN-based smoothing (see Section \ref{MDN-method}) provides a reliable estimate of the LOSVD and its velocity moments prior to the Schwarzschild modeling. Figure~\ref{MDN} shows the second- and fourth-order velocity moment profiles for two representative mock data sets (Mock 1 and Mock 2). The gray points show the moments measured directly from the mock data, with their uncertainties estimated by bootstrap resampling. The black points show the moments obtained from the MDN, while the blue dashed curves indicate the true profiles used to generate the mock data. The upper and lower rows show Mock 1 and Mock 2, respectively, representing cases in which the true second moment decreases and increases toward the center.

The MDN recovers the overall profile of the second moment, with the true values falling within the estimated $1\sigma$ uncertainties across almost all spatial bins. Although the $1\sigma$ error bars naturally broaden toward the inner and outer regions due to the smaller sample size, the model properly reflects the increased uncertainty in these data-limited regimes.

The fourth moment is intrinsically more difficult to recover accurately because it scales as $v^4$ and is therefore highly sensitive to the tails of the velocity distribution. As shown in the right panels of Figure~\ref{MDN}, the MDN recovers a smooth profile that follows the overall radial trend. However, the true values occasionally fall outside the estimated \(1\sigma\) bounds. This contrasts with the second moment, for which the true values are almost entirely enclosed by these bounds. To prevent the uncertain fourth-moment estimates from over-constraining the model or introducing spurious biases, we use a Mahalanobis distance based on their joint covariance matrix in the subsequent Schwarzschild modeling (see Section \ref{section_objective_function}). This approach accounts for correlations between the two moments and their different physical scales. It therefore allows their information content and associated uncertainties to be propagated consistently into the likelihood.

In addition, smoothing the moment profiles introduces correlations between neighboring radial bins. We therefore account for these correlations by applying a temperature scaling to the effective $\chi^2$ before MCMC sampling (see Section~\ref{likelihood_method}). This prevents the smoothed moment profiles from being treated as fully independent measurements.

\subsection{The Recovery of Density Profile}
\begin{figure*}
    \centering
    \includegraphics[width=\textwidth]{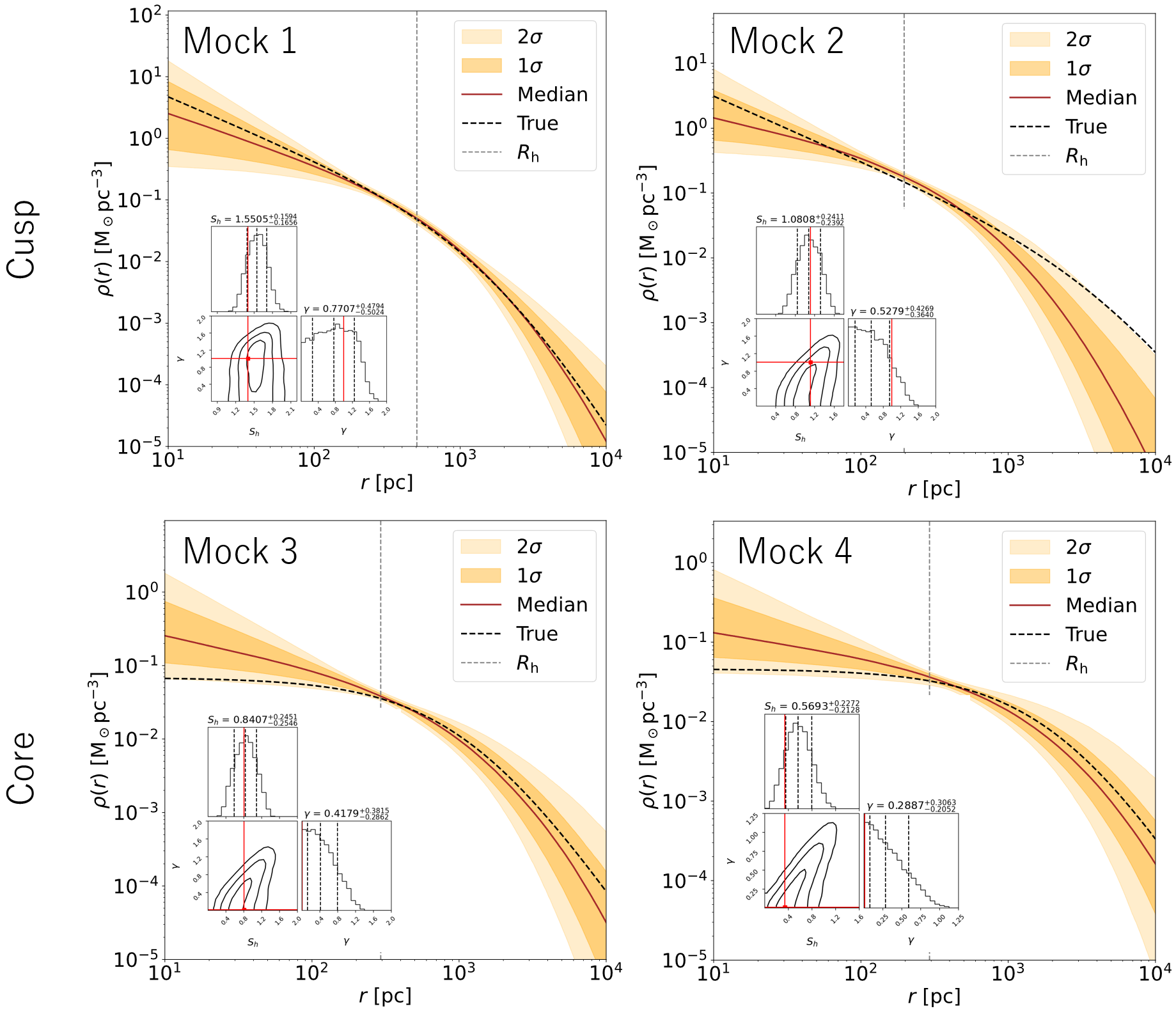}
    \caption{Recovery of the dark matter density profiles for four mock halos. The upper and lower rows show cusp and core models, respectively. The orange shaded regions indicate the $1\sigma$ and $2\sigma$ posterior credible intervals, and the red solid lines show the posterior medians. The black dashed lines indicate the true density profiles used to generate the mock data. The vertical gray dashed lines mark the projected half-light radius $R_h$. The inset panels show the posterior distributions of the logarithmic density slope at the half-light radius, $S_h$, and the asymptotic inner slope, $\gamma$. The two-dimensional contours are shown up to the $3\sigma$ level. In the one-dimensional posteriors, the black dashed lines indicate the posterior median and $1\sigma$ credible interval, while the red solid lines indicate the true values. 
    }
    \label{Mock_density}
\end{figure*}

We test whether our dynamical model can recover the dark matter density profiles of the mock data sets using the same modeling framework and analysis procedure as adopted for the observed galaxies in the main text (see Section \ref{method}). For all mock analyses, we adopt the uniform prior on $S_h$ (Prior 1).  Figure~\ref{Mock_density} shows the posterior density profiles and the posterior distributions of $S_h$ and $\gamma$ for four mock halos. The dark and light orange regions indicate the $1\sigma$ and $2\sigma$ credible intervals, respectively, and the red solid lines show the posterior medians. The true density profiles are shown by the black dashed lines. For the two-dimensional posterior distributions of $S_h$ and $\gamma$, contours are shown up to the $3\sigma$ level. The one-dimensional posterior distributions show the posterior medians and $1\sigma$ credible intervals with black dashed lines, while the true values are indicated by the red solid lines.

For all four mock halos, the true density profile is broadly recovered within the $2\sigma$ credible interval. The recovery is particularly good for the cusp models (Mock 1 and Mock 2) in the inner regions, where the true profiles are generally contained within the $1\sigma$ credible interval. For the core models (Mock 3 and Mock 4), the true profiles lie closer to the edge of the $2\sigma$ region in the innermost part of the profile. One possible reason for this behavior is the lower bound $\gamma>0$ adopted in our MCMC sampling. Since the parameter space does not extend to $\gamma<0$, the posterior volume near $\gamma=0$ can be asymmetric, which may affect the recovery of shallow central profiles. Although recent studies have allowed $\gamma<0$ (e.g., \citealt{2024ApJ...970....1V,2026arXiv260626218N}), we restrict our analysis to profiles whose central density is constant or increases inward. As discussed in Section~\ref{first_discussion}, the stellar kinematic data constrain the asymptotic inner slope $\gamma$ only weakly, whereas the finite-radius slope around the half-light radius, $S_h$, is much better constrained. We also verified that extending the prior to $\gamma<0$ changes the inferred $S_h$ only marginally, indicating that our main conclusions regarding $S_h$ are insensitive to the lower bound on $\gamma$.

A more notable deviation is seen in Mock 2. Although the overall density profile is still recovered within approximately the $2\sigma$ credible region, the posterior distribution of $\gamma$ peaks away from its true value, and the median density profile shows a noticeable deviation from the true profile. We attribute this behavior to the choice of a uniform prior on $S_h$ (Prior 1). As discussed in Section \ref{first_discussion}, Prior 1 suppresses large values of $r_0$; in particular, the prior approaches $p(r_0)\propto r_0^{-2}$ in the large-$r_0$ regime (see equation \eqref{jacobian_1}). Mock 2 has a relatively large scale radius, $r_0=25R_h$, and is therefore strongly affected by this prior. Such a large $r_0$ corresponds to a density profile whose logarithmic slope changes only gradually over the radial range probed by the data. The suppression of large $r_0$ therefore makes these profiles less favored by the adopted prior, which in turn shifts the posterior distribution of $\gamma$ away from the true value.

This test indicates that the combination of our model and the $S_h$-uniform prior is less effective at recovering halos with very large scale radii. This limitation is important when interpreting the asymptotic parameters of the density profile. On the other hand, the recovery of $S_h$ is good for all four mock halos, both in terms of the posterior distribution and the size of its uncertainty. These results support the robustness of the $S_h$ constraint, while emphasizing that the recovery of $\gamma$ and the detailed density profile can become sensitive to the adopted prior, particularly for halos with large $r_0$.

\subsection{The Recovery of Anisotropy Profile}
\begin{figure*}
    \centering
    \includegraphics[width=\textwidth]{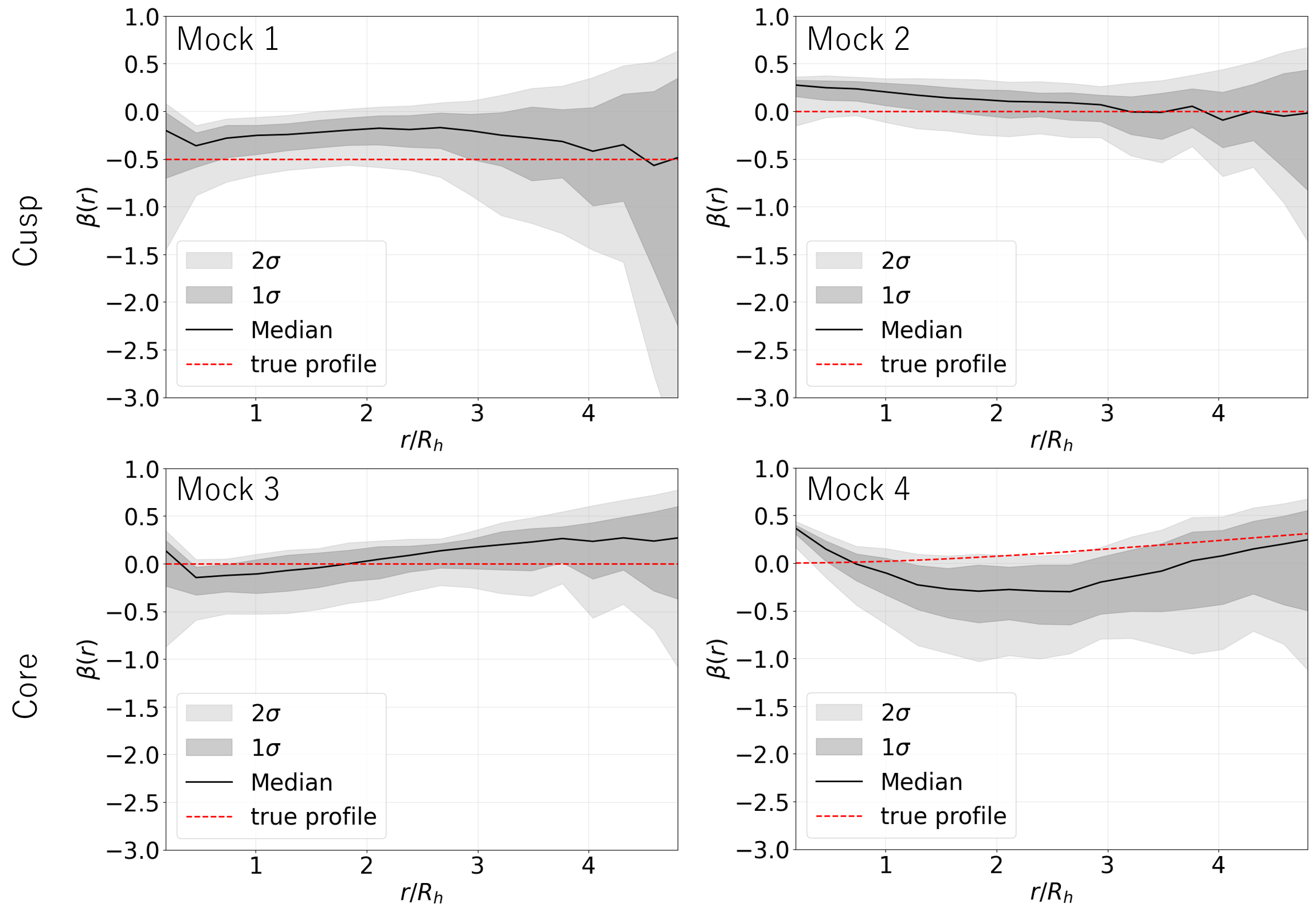}
    \caption{Recovery of the velocity anisotropy profiles for the four mock halos. The upper and lower rows show cusp and core models, respectively. The black solid lines show the posterior median of $\beta(r)$, while the dark and light gray regions indicate the $1\sigma$ and $2\sigma$ credible intervals, respectively. The red dashed lines show the true anisotropy profiles used to generate the mock data. The radius is normalized by the projected half-light radius $R_h$.
    }
    \label{Mock_beta}
\end{figure*}

Testing the recovery of the velocity anisotropy is also important for assessing the capability of our Schwarzschild model. Figure~\ref{Mock_beta} shows the posterior velocity anisotropy profiles obtained for the four mock data sets.  The true anisotropy profiles are recovered within, or close to, the $2\sigma$ credible intervals across all four mock halos.

A systematic offset is most apparent in the inner regions. For the cusp models (Mock 1 and Mock 2), the posterior median tends to be shifted toward more radial anisotropy relative to the true profile, whereas for the core models (Mock 3 and Mock 4), it tends to be shifted toward more tangential anisotropy. These deviations are not necessarily confined to the central region, however, and their radial behavior differs among the mock data sets. The origin of this behavior is not fully understood, but it may reflect the density--anisotropy degeneracy inherent in dynamical modeling: changes in the inferred mass profile can be partially compensated by changes in the orbital anisotropy while producing similar projected kinematic observables. 

The uncertainty in $\beta(r)$ generally increases toward larger radii, as the number of tracer stars decreases, leaving less kinematic information to constrain the orbital structure of the outer halo. However, the median anisotropy profiles approach the true values toward larger radii in three of the four mock halos (Mock 1, Mock 2, Mock 4). This trend may be related to a weakening of the slope--anisotropy degeneracy in the outer regions, as discussed in Section~\ref{second_discussion}, although it is not evident in all mock cases.

These mock tests indicate that our model can capture the overall trend of the anisotropy profile, although the detailed shape of $\beta(r)$ can be subject to systematic deviations from the true profile. The offsets seen in the mock tests are therefore particularly relevant when interpreting the velocity anisotropy inferred for the observed dwarf galaxies, and motivate some caution in assigning physical significance to detailed features of the recovered profiles.


\bibliography{sample701}
\bibliographystyle{aasjournalv7.1}



\end{document}